\pdfoutput=1
\documentclass[journal=jacsat,manuscript=review]{achemso}
\setkeys{acs}{maxauthors = 0}   %

\usepackage[version=3]{mhchem} %
\usepackage{physics}
\usepackage{subcaption}
\usepackage{hyperref}
\usepackage{color}
\usepackage{graphicx}
\usepackage{ifthen}
\usepackage{booktabs}  %
\usepackage[utf8]{inputenc}
\usepackage[normalem]{ulem} %

\ifdefined\DeclareUnicodeCharacter
  \DeclareUnicodeCharacter{1F534}{\textbf{[!]}}
  \DeclareUnicodeCharacter{1F511}{\textbf{[key]}}
  \DeclareUnicodeCharacter{26A0}{\textbf{[!]}}
  \DeclareUnicodeCharacter{FE0F}{}
\fi

\usepackage{amsmath}
\usepackage{textcomp} %
\usepackage{placeins} %

\newcommand{\ai}[1]{{\color{red}[AI: #1]}}  %
\makeatletter
\renewcommand*\acs@author@list@main{%
  \advance\@tempcnta\@ne\relax
  \def\acs@author@footnotes{}%
  \ifnum\@tempcnta=5\relax\space and\else\acs@author@list@and\fi
  \ifnum\@tempcnta=\tw@\relax\else\ifnum\@tempcnta=6\relax\else\space\fi\fi
  \@nameuse{@author@\@roman\@tempcnta}%
  \acs@author@list@comma
  \acs@author@star
  \acs@author@affil
  \acs@author@affil@also
  \acs@author@affil@alt
  \ifx\@empty\acs@author@footnotes\else
    \textsuperscript{\acs@author@footnotes}%
  \fi
  \ifnum\@tempcnta=\@ne\relax\ifnum\acs@author@cnt>\@ne\relax\\\fi\fi
  \ifnum\@tempcnta=5\relax\ifnum\acs@author@cnt>5\relax\\\fi\fi
  \ifnum\@tempcnta<\acs@author@cnt\relax
    \expandafter\acs@author@list@main
  \fi
}
\makeatother
\newcommand{\pd}[1]{\textcolor{blue}{[#1]}}     %

\DeclareRobustCommand{\revdel}[1]{{\color{red}\sout{#1}}}  %

\newcommand{\dgbookmarks}{on}

\makeatletter
\ifthenelse{\equal{\dgbookmarks}{on}}{%
  \newcounter{dgbm}
  \let\dg@sect\section
  \let\dg@subsect\subsection
  \newcommand{\dg@mark}[2]{\refstepcounter{dgbm}\pdfbookmark[#1]{#2}{dgbm.\thedgbm}}
  \renewcommand{\section}{\@ifstar{\dg@sectstar}{\@ifnextchar[{\dg@sectopt}{\dg@sectplain}}}
  \def\dg@sectstar#1{\dg@mark{1}{#1}\dg@sect*{#1}}
  \def\dg@sectopt[#1]#2{\dg@mark{1}{#1}\dg@sect[#1]{#2}}
  \def\dg@sectplain#1{\dg@mark{1}{#1}\dg@sect{#1}}
  \renewcommand{\subsection}{\@ifstar{\dg@substar}{\@ifnextchar[{\dg@subopt}{\dg@subplain}}}
  \def\dg@substar#1{\dg@mark{2}{#1}\dg@subsect*{#1}}
  \def\dg@subopt[#1]#2{\dg@mark{2}{#1}\dg@subsect[#1]{#2}}
  \def\dg@subplain#1{\dg@mark{2}{#1}\dg@subsect{#1}}
}{}
\makeatother

\newcommand{\docstatus}{preprint}          %

\newcommand{\logostrip}{%
  \raisebox{-0.5\height}{\includegraphics[height=0.85in]{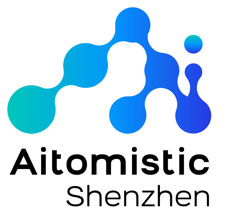}}\hspace{1.6em}%
  \raisebox{-0.5\height}{\includegraphics[height=0.95in]{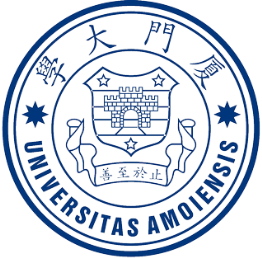}}\hspace{1.6em}%
  \raisebox{-0.5\height}{\includegraphics[height=0.80in]{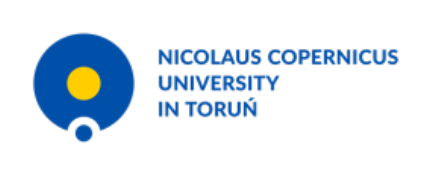}}%
}
\newcommand{\draftbanner}{}

\newcommand{\statusblock}{%
  \ifthenelse{\equal{\docstatus}{submission}}{}{%
    \begin{center}\logostrip\end{center}\vspace{0.3em}%
    \ifthenelse{\equal{\docstatus}{draft}}{\draftbanner}{}%
  }%
}

\author{Pavlo O. Dral}
\email{dral@xmu.edu.cn}
\affiliation[XMU]{State Key Laboratory of Physical Chemistry of Solid Surfaces, College of Chemistry and Chemical Engineering, and Fujian Provincial Key Laboratory of Theoretical and Computational Chemistry, Xiamen University, Xiamen, Fujian 361005, China}
\alsoaffiliation[UMK]{Institute of Physics, Faculty of Physics, Astronomy, and Informatics, Nicolaus Copernicus University in Toru\'n, ul. Grudziadzka 5, 87-100 Toru\'n, Poland}
\alsoaffiliation[UMK2]{Institute of Advanced Studies, Nicolaus Copernicus University in Toruń, ul. Wileńska 4, 87-100 Toruń, Poland}
\alsoaffiliation[Aitomistic]{Aitomistic, Shenzhen 518000, China}
\author{Yi-Fan Hou}
\affiliation[XMU]{State Key Laboratory of Physical Chemistry of Solid Surfaces, College of Chemistry and Chemical Engineering, and Fujian Provincial Key Laboratory of Theoretical and Computational Chemistry, Xiamen University, Xiamen, Fujian 361005, China}
\author{Yuxinxin Chen}
\affiliation[XMU]{State Key Laboratory of Physical Chemistry of Solid Surfaces, College of Chemistry and Chemical Engineering, and Fujian Provincial Key Laboratory of Theoretical and Computational Chemistry, Xiamen University, Xiamen, Fujian 361005, China}
\author{Hassan Nawaz}
\affiliation[XMU]{State Key Laboratory of Physical Chemistry of Solid Surfaces, College of Chemistry and Chemical Engineering, and Fujian Provincial Key Laboratory of Theoretical and Computational Chemistry, Xiamen University, Xiamen, Fujian 361005, China}
\author{Mateusz Witkowski}
\affiliation[UMK]{Institute of Physics, Faculty of Physics, Astronomy, and Informatics, Nicolaus Copernicus University in Toru\'n, ul. Grudziadzka 5, 87-100 Toru\'n, Poland}
\alsoaffiliation[UMK2]{Institute of Advanced Studies, Nicolaus Copernicus University in Toruń, ul. Wileńska 4, 87-100 Toruń, Poland}
\author{Szymon Śmiga}
\affiliation[UMK]{Institute of Physics, Faculty of Physics, Astronomy, and Informatics, Nicolaus Copernicus University in Toru\'n, ul. Grudziadzka 5, 87-100 Toru\'n, Poland}
\alsoaffiliation[UMK2]{Institute of Advanced Studies, Nicolaus Copernicus University in Toruń, ul. Wileńska 4, 87-100 Toruń, Poland}
\author{Ireneusz Grabowski}
\affiliation[UMK]{Institute of Physics, Faculty of Physics, Astronomy, and Informatics, Nicolaus Copernicus University in Toru\'n, ul. Grudziadzka 5, 87-100 Toru\'n, Poland}
\alsoaffiliation[UMK2]{Institute of Advanced Studies, Nicolaus Copernicus University in Toruń, ul. Wileńska 4, 87-100 Toruń, Poland}

\title[Can MLPs Be More Accurate Than the Methods They Are Trained On?]
  {Can Machine Learning Potentials Be More Accurate Than the Methods They Are Trained On?}

\makeatletter

\newcommand*\l@sisection{\@dottedtocline{1}{0em}{0em}}
\newcommand*\l@sisubsection{\@dottedtocline{2}{1.5em}{0em}}

\newcommand{\sicontentsname}{Table of Contents}

\newcommand{\sisection}[1]{%
  \phantomsection\addcontentsline{sit}{sisection}{#1}\section*{#1}}
\newcommand{\sisubsection}[1]{%
  \phantomsection\addcontentsline{sit}{sisubsection}{#1}\subsection*{#1}}

\newcommand{\printsicontents}{%
  \begingroup
    \c@tocdepth=2\relax
    \renewcommand*\@pnumwidth{1.8em}%
    \renewcommand*\@tocrmarg{2.4em}%
    \section*{\sicontentsname}%
    \@starttoc{sit}%
  \endgroup}

\makeatother

\mciteErrorOnUnknownfalse
\makeatletter
\newwrite\si@auxout
\newcommand{\startsireferences}[1]{%
  \if@filesw
    \immediate\openout\si@auxout=\jobname-si.aux\relax
    \immediate\write\si@auxout{\relax}%
    \immediate\write\si@auxout{\string\citation{achemso-control}}%
    \immediate\write\si@auxout{\string\bibstyle{\acs@bibstyle}}%
    \immediate\write\si@auxout{\string\bibdata{acs-\jobname,#1}}%
  \fi
  \gdef\mciteauxout{\si@auxout}%
  \gdef\mcitetrackID{si}%
  \gdef\@extra@b@citeb{@SI}%
  \global\let\si@std@citex\@citex
  \gdef\@citex[##1][##2]##3{%
    \begingroup\let\@auxout\si@auxout
    \si@std@citex[##1][##2]{##3}%
    \endgroup}%
  \global\let\si@main@citenumfont\citenumfont
  \gdef\citenumfont##1{\si@main@citenumfont{S##1}}%
}
\newcommand{\printsireferences}{%
  \begingroup
    \def\mcitebibtrackID{si}%
    \let\si@main@bibnumfmt\bibnumfmt
    \def\bibnumfmt##1{\si@main@bibnumfmt{S##1}}%
    \def\bibsection{\sisection{References}}%
    \let\si@std@NAT@wrout\NAT@wrout
    \def\NAT@wrout##1##2##3##4##5{\si@std@NAT@wrout{##1}{##2}{##3}{##4}{##5@SI}}%
    \@input@{beyond-training-si.bbl}%
  \endgroup
  \if@filesw\immediate\closeout\si@auxout\fi
}
\makeatother

\begin{document}

\statusblock

\clearpage
\begin{abstract}
Although machine learning potentials (MLPs) generalize across configurational and chemical compound spaces,
their accuracy is commonly believed to be limited by their training methods, with improvements beyond this
limit achieved so far only by refitting to experimental observables. In this Perspective, we argue that
MLPs can also be made more accurate than their training methods non-empirically by generalizing across a
third space, that of quantum-chemical methods. We propose routes to method-transferable MLPs that predict
at unseen methods, with working implementations in both density-functional and wavefunction-based theories
showing that they can be more accurate than any of their training methods. Aiming beyond the accuracy of the training methods could also make universal MLPs more accurate than single-method models, even when the method-transferable MLPs remain less accurate than those QM methods. Method-transferable MLPs will shape the construction of training datasets and the evaluation of models, pushing the accuracy limits of chemical simulations.
\end{abstract}

\newcommand{\tier}[1]{%
  \ifthenelse{\equal{\docstatus}{draft}}{%
    \marginpar{\footnotesize\textcolor{%
      \ifthenelse{\equal{#1}{spine}}{blue}{gray}}{\textsf{#1}}}%
  }{}%
}

\section{Introduction}
\label{sec:intro}

\begin{figure}[htbp]
\centering
\includegraphics[width=\linewidth,trim=0 0.16in 0 0,clip]{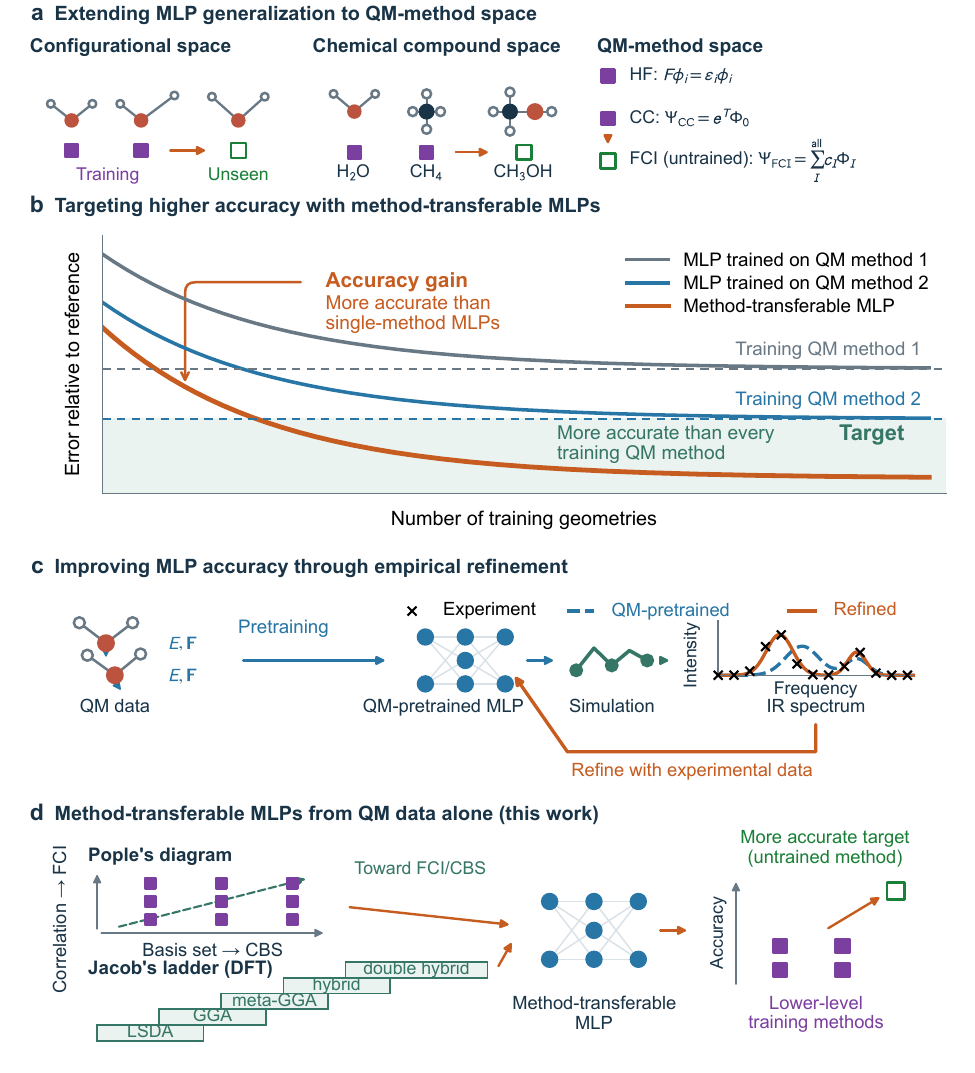}
\caption{\textbf{Exceeding the accuracy of the training methods with MLPs.} (a) Generalization across configurations, compounds and QM methods. FCI includes all determinants (all excitation ranks) in the chosen basis. Throughout this Perspective, filled/open markers denote training/unseen cases; noncircular method symbols distinguish them from atoms and neurons. (b) Schematic MLP error against the same more accurate reference as the number of training geometries grows. The dashed levels are limits of exact reproduction of the training methods. The orange arrow marks the accuracy gain; green shading indicates the target: accuracy beyond every training QM method. Even when this target is not reached, method-transferable MLPs can improve on single-method MLPs. (c) Empirical refinement compares simulated observables, such as IR spectra, with experiment to refine a QM-pretrained MLP.\cite{Roecken2024_AlternatingDensityFunctional,Gumber2025_RefiningPotentialsUranium,Han2025_RefiningWaterPotential} (d) Known relationships between QM methods, represented by Pople's diagram and Jacob's ladder, could guide method-transferable MLPs.}
\label{fig:concept}
\end{figure}

Machine learning potentials (MLPs) are actively reshaping how atomistic simulations are
performed.\cite{Xia2025_EvolutionMachineLearning,Jinnouchi2025_MachineLearningForce,Kulichenko2024_DataGenerationMachine,Behler2021_FourGenerationsHigh,Perspective-ML-in-QC}
MLPs, with their superior computational speed compared to quantum mechanical (QM) methods, enable
simulations at unprecedented time scales and system sizes, allowing for the gain of previously unattainable
insight into processes in molecules and
materials.\cite{Bocus2024_OperandoNatureIsobutene,Shi2022_MethanolSynthesisCo,Yan2024_AcceleratingReliableMultiscale,Nandi2021_QuasiclassicalSimulationsBased,Friederich2021_MachineLearnedPotentials,Deringer2021_OriginsStructuralElectronic,Westermayr2022_DeepLearningStudy,Wang2024_MachineLearningPhotodynamics}
Their usefulness rests
on generalization across two spaces. In configurational space, a model learns the mapping from
nuclear geometry to potential energy and predicts energies for geometries it has never
seen.\cite{Behler2007_GeneralizedNeuralNetwork} In chemical compound space, once the total energy is
split into atomic contributions, a model trained on some molecules predicts energies for unseen
compounds (Figure~\ref{fig:concept}a).\cite{Smith2019_ApproachingCoupledCluster,Takamoto2022_TowardsUniversalNeural}

\FloatBarrier

No such generalization exists for the method space, where MLP would make predictions for unseen QM methods, fundamentally limiting the accuracy of an MLP to that of the training method used to generate the training data (Figure~\ref{fig:concept}b). No surprise, then, that the belief that an MLP cannot surpass the accuracy of its training method is widely held and expressed in authoritative reviews and books,\cite{Behler2015_ConstructingHighDimensional,Behler2016_PerspectiveMachineLearning,Hellstroem2020_HighDimensionalNeural,Omranpour2024_PerspectiveAtomisticSimulations}
including ours.\cite{Perspective-ML-in-QC,ML-ES,OMNIP2x}

Inspiringly, MLPs have been made more accurate than their training methods, but, unfortunately, so far only empirically. MLPs empirically refined on experimental observables
predict those observables more accurately than the QM methods they were initially trained on (Figure~\ref{fig:concept}c).\cite{Roecken2024_AlternatingDensityFunctional,Han2025_RefiningWaterPotential,Gumber2025_RefiningPotentialsUranium}
Spectroscopic refinement of potential-energy surfaces against measured rovibrational levels has done
this for decades.\cite{Huang2012_Ames1CO2} The limits of this route are structural: it requires experiments, which are often unavailable and
carry their own errors. Most importantly, it offers no systematic way to improve an MLP from theory alone.
Quantum chemistry has systematic orderings of its own methods, such as Jacob's ladder for density
functionals\cite{JohnPPerdew2001_JacobSLadder} and the Pople diagram for the treatment of correlation and
the basis set.\cite{Pople1965_DimensionalChartQuantum} Methods therefore stand in known relations to
one another, and a model that learns those relations could be made transferable across them.

In this Perspective, we argue that MLPs can exceed the accuracy of their training methods
non-empirically, from QM data alone (Figure~\ref{fig:concept}d). Just as an MLP can be made transferable in
configurational and chemical compound space, it can be made transferable in a third space, that of
QM methods. We envision and demonstrate possible constructions
of such method-transferable MLPs, predicting at unseen methods.
The constructions we propose below extend existing approaches to
learning across methods, which predict only at their training methods (reviewed in the next section).

\FloatBarrier

\section{Existing MLPs learning across methods}
\label{sec:map}

Existing approaches to learn across methods were designed with the motivation to exploit the heterogeneous data obtained with different levels, with the major goal of reducing the required number of expensive high-level QM calculations, rather than improving upon the training QM methods. Plenty of such methods were developed over time.\cite{PavloODral2023_LearningMultipleQuantum}
Because these methods are central to the development of method-transferable MLPs, here, we give a brief overview of such methods.

\begin{figure}[htbp]
\centering
\includegraphics[width=\linewidth]{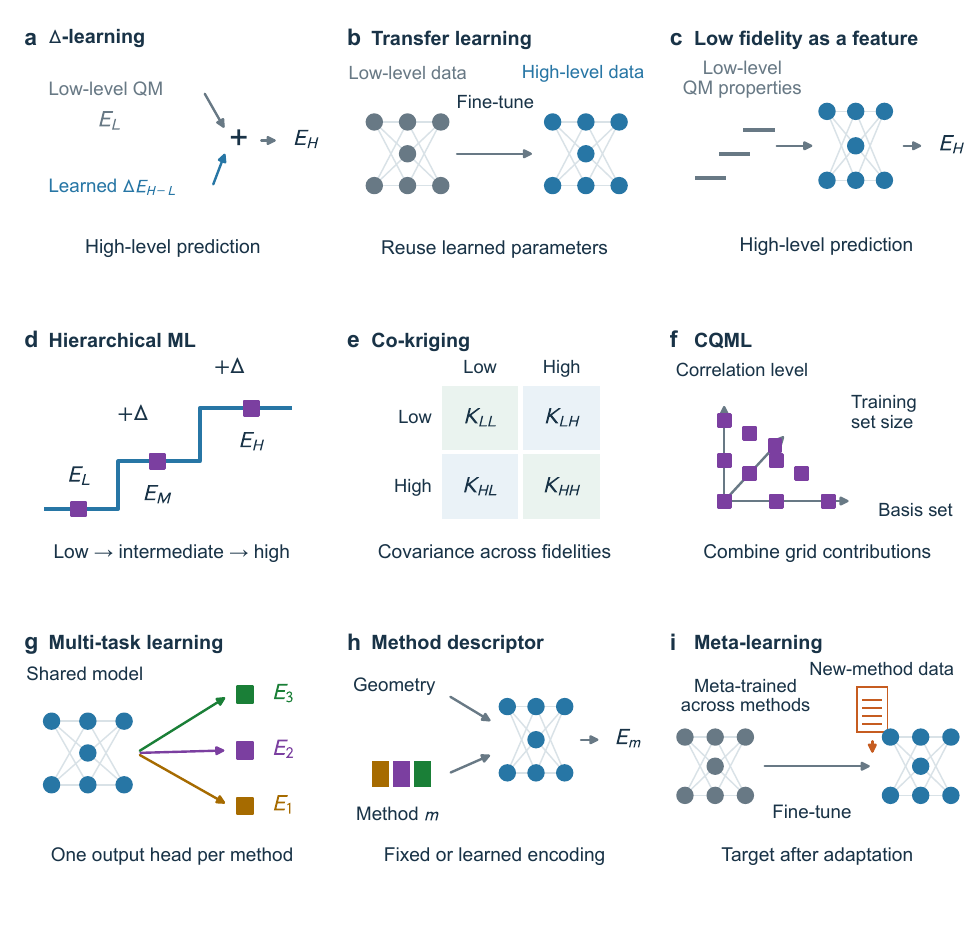}
\caption{\textbf{Existing approaches for learning in method space.} $E$ denotes energy; $L$, $M$ and $H$ denote low, intermediate and high levels. $\Delta E_{H-L}$ is a learned high-minus-low energy correction; numbered subscripts and $m$ identify individual QM methods. (a) $\Delta$-learning; (b) transfer learning; (c) low-fidelity properties as features; (d) hierarchical machine learning; (e) co-kriging, where $K_{LL}$ and $K_{HH}$ are low- and high-fidelity covariance blocks and the off-diagonal blocks couple them; (f) the combination technique CQML, whose third axis is training-set size; (g) multi-task outputs; (h) fixed or learned method descriptors as input to MLP; and (i) meta-learning followed by adaptation using new-method training data.}
\label{fig:map}
\end{figure}

Two of the most common approaches---$\Delta$-learning~\cite{DeltaLearning} (Figure~\ref{fig:map}a) and transfer learning~\cite{Pan2010_SurveyTransferLearning} (Figure~\ref{fig:map}b)---require a low- and high-level set of training QM methods to generate the reference data.  $\Delta$-learning is simply learning the difference between them, and requires the low-level QM method to perform the baseline calculations at inference time. Transfer learning first trains an MLP on the source, low-level QM data and then fine-tunes it to the target, high-level QM data to obtain two sets of MLPs (with the intention to only use the one trained on a high level) as, e.g., has been done in ANI-1ccx~\cite{Smith2019_ApproachingCoupledCluster} and other potentials~\cite{Kser2021_TransferLearningCcsd}. None of the transfer-learned models requires baseline QM calculations at inference time, but without a physical baseline, the calculations might produce unphysical results. The two approaches can be efficiently combined such that the $\Delta$-learning correction is trained in two stages like in transfer learning: this was utilized, e.g., in our AIQM methods\cite{AIQM1,AIQM2} which correct the semi-empirical QM baseline with an MLP-like correction to the coupled-cluster level. The MLP-like correction was pretrained on DFT-level corrections before it was fine-tuned to yield the coupled-cluster-level corrections.

Variants and extensions of these approaches exist. Low fidelity as a feature (Figure~\ref{fig:map}c) utilizes the baseline QM properties as an input, and has been used on dopant formation energies in
hafnia\cite{Batra2019_MultifidelityInformationFusion}---a similar approach combined with $\Delta$-learning has been adopted in molecular-orbital-based\cite{Welborn2018_TransferabilityMachineLearning} potentials such as OrbNet Denali,\cite{Qiao2020_OrbnetDeepLearning,Christensen2021_OrbnetDenaliMachine} etc. Hierarchical machine learning
stacks $\Delta$-models to systematically improve across the hierarchy of QM methods (Figure~\ref{fig:map}d), which enabled very data-efficient training reducing the cost of generating high-level reference data by 99\% and enabled obtaining spectroscopic accuracy (an error of ca.\
1~cm$^{-1}$),\cite{hML} and was later adopted in other applications.\cite{Nandi2021_MachineLearningPotential,Bowman2023_DeltaMachineLearned}

Embedding correlation between methods is done with methods such as co-kriging (Figure~\ref{fig:map}e) that exploits the covariance between fidelities in a kernel, and it extends, in principle, to
more than two levels.~\cite{Pilania2017_MultiFidelityMachine,Batra2019_MultifidelityInformationFusion} The combination technique CQML (Figure~\ref{fig:map}f) places the
levels on a grid over correlation treatment, basis set, and training set size, its third axis
being the number of training molecules, and this line of
work has since been extended further and optimized.\cite{Zaspel2019_BoostingQuantumMachine,VivinVinod2023_MultifidelityMachineLearning,Heinen2024_ReducingTrainingData,Vinod2024_OptimizedMultifidelity,Vinod2025_InvestigatingDataHierarchies}
All of them are built to deliver the top of the hierarchy they were given.

A new route to learn heterogeneous data, which we argue is poised to be the new standard one, is learning all training methods with MLPs in a single model. Then the QM method can be chosen at inference, so that predictions targeting any training QM method stay available. One way to achieve this is a separate
output head in a neural network (NN) per method with multi-task learning (Figure~\ref{fig:map}g).~\cite{Jacobson2023_LeveragingMultitaskLearning,Messerly2025_MultiFidelityLearning,Zhang2024_DpaLargeAtomic,IlyesBatatia2025_CrossLearningBetween}
Another way is, arguably, more elegant: providing a (QM) method descriptor as an input to the network (Figure~\ref{fig:map}h) such as in PFP,~\cite{Takamoto2022_TowardsUniversalNeural} SevenNet-MF,\cite{JaesunKim2024_DataEfficientMultifidelity}  and our own all-in-one model OMNI-P1.\cite{AIO-OMNI-P1}
It is the method we now advocate and extend later in this Perspective. DPA-3 uses a related approach with a dataset descriptor as input.\cite{DuoZhang2025_GraphNeuralNetwork}

The input can also be learned rather than fixed.
Oerder et al.\ train one embedding for each of ten reference methods drawn from
MultiXC-QM9,\cite{Oerder2025_MultiFidelityLearning,Nandi2023_MultixcQmLarge} and Chen et al.\ embed
the fidelity as a trainable vector alongside the elemental embeddings and let measured band gaps
enter as one more fidelity,\cite{Chen2021_MultifidelityMaterials} which puts experiment in as one
more trained label rather than as a first-principles, systematic approach to improve past the training methods.

Meta-learning is another relevant approach (Figure~\ref{fig:map}i), where it is used to obtain a representation that refits cheaply to a new method.\cite{Allen2024_LearningTogether} The model still has to be
fine-tuned before it predicts at any method at all.

None of these approaches predicts at a method that was not in its training set, and we pinpoint this as the major limitation that caps the accuracy of MLPs.

\FloatBarrier

\section{Exceeding the accuracy of the training methods}
\label{sec:breaking}

Given the huge body of studies around MLPs and learning across method space, it is surprising that the field seems to concede to the `impossibility' of training MLPs more accurate than the training QM methods. Hence, here we conceive several ideas to challenge this assumption and inspire a new era in MLP development informed by the systematic relationships known in QM theory. Beyond ideas, we provide architectural recipes and numerical demonstrations of the feasibility of such an approach.

\subsection{Taking a page from the quantum chemistry book: repurposing existing extrapolation schemes for MLPs}
\label{sec:extrapolation}

Think about established extrapolation schemes in QM theory such as Helgaker's complete basis set (CBS)
extrapolation scheme\cite{Helgaker1997_BasisConvergenceCorrelated,Halkier1998_BasisSetConvergence} or focal-point
analysis\cite{Csaszar1998_PursuitAbInitioLimit}. While these methods enable predictions beyond the seen methods, their practical disadvantage is the need to perform additional QM calculations with non-negligible cost.

The reader can now easily connect the dots, and see that many of the existing MLPs learning across method space, such as variants based on multi-task learning or on using a method descriptor as input, can readily be repurposed to produce outputs at the same (training) QM methods that enter established extrapolation schemes but at negligible cost. This is the \emph{extrapolation route}. Until now, the studies extrapolated first and only then trained MLP on already extrapolated values, by transfer learning as ANI-1ccx\cite{Smith2019_ApproachingCoupledCluster} or by stacking $\Delta$-models up to the extrapolated level as in hierarchical ML,\cite{hML} so that the extrapolation scheme is fixed before the training data are generated, and the extrapolated values exist only for the geometries at which every QM method of the scheme was calculated. We propose to reverse this order and decouple training and extrapolation, which brings practical advantages: one MLP returns every training QM method, which can be used in various extrapolation schemes as a simple post-processing step (i.e., extrapolation scheme does not need to be pre-defined before training MLP), different geometries can be labeled with different methods, e.g., potentially improving such workflows as active learning, where progressively fewer geometries can be labeled with the more expensive QM methods.

\begin{figure}[htbp]
\centering
\includegraphics[width=0.985\linewidth]{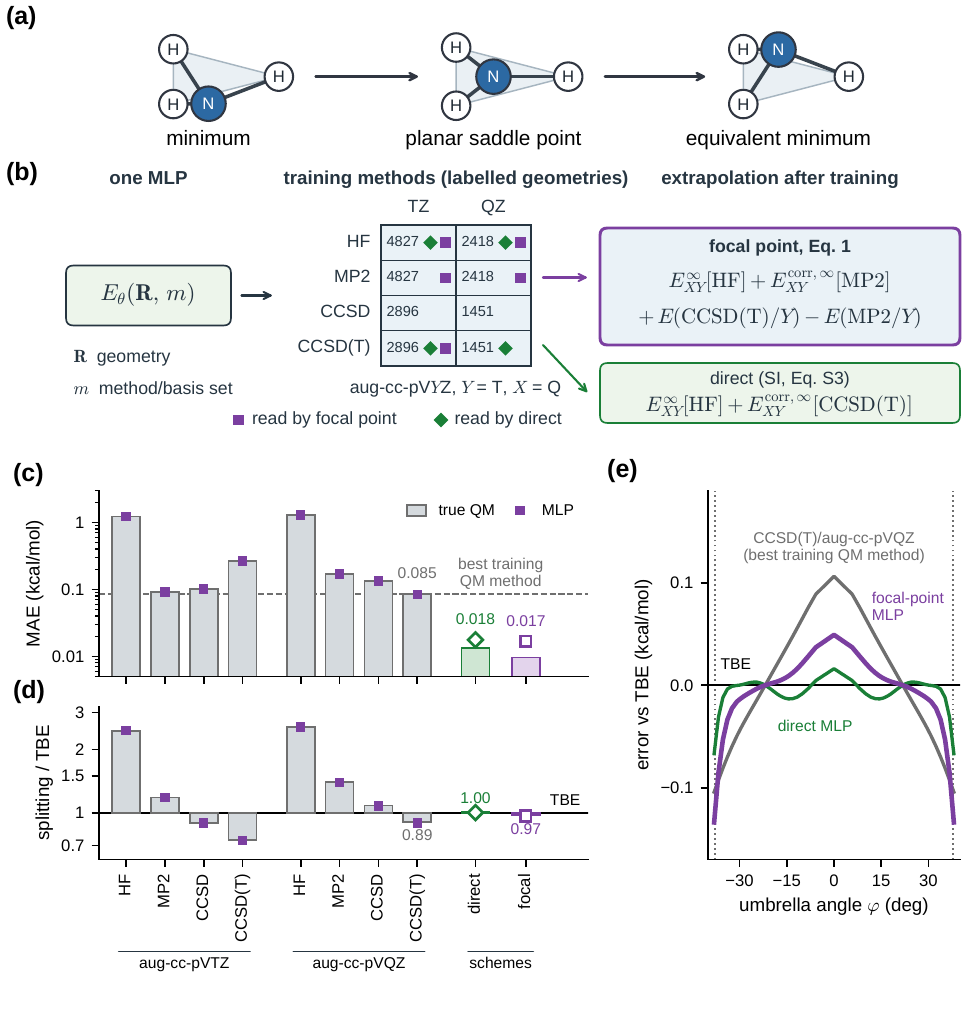}%
\caption{\textbf{One MLP, extrapolation after training: the inversion of ammonia.} (a) Inversion through the planar saddle point. (b) One MLP predicts energies at geometry $\mathbf{R}$ for every training QM method $m$ (method and basis set). Numbers count labelled training geometries used for training; purple squares and green diamonds identify energies entering the focal-point (Eq.~\ref{eq:schemeF}) and direct-extrapolation (SI, Eq.~\ref{eq:schemeD}) schemes, respectively; $Y$ = aug-cc-pVTZ, $X$ = aug-cc-pVQZ. (c) Mean absolute error (MAE) of relative energies against the theoretical best estimate (TBE) on the held-out test set, for all training methods and both schemes. Bars: true QM; symbols: corresponding MLP predictions (filled: training methods; open: unseen schemes); dashed line: best training QM method. (d) One-dimensional inversion splitting on each relaxed surface, divided by the TBE-surface splitting; bars and symbols as in (c). (e) Relaxed inversion-profile errors against the TBE for both MLP schemes and the best training QM method in true QM. The umbrella angle $\varphi$ measures N--H tilt out of the hydrogen plane; N--H distances are relaxed on each surface. Dotted lines bound the training geometries.}
\label{fig:extrapolation-overview}
\end{figure}

As a demonstration, we construct an MLP that describes the umbrella inversion of ammonia (Figure~\ref{fig:extrapolation-overview}a) more accurately than any of its training QM methods. Ammonia inversion was one of the conformational energy prototypes in the original focal-point analysis studies,\cite{Csaszar1998_PursuitAbInitioLimit} and we use this scheme to target CCSD(T) at the CBS limit [CCSD(T)/CBS] from calculations with two basis sets with the cardinal numbers $X > Y$. It extrapolates the HF energy and the correlation energy of the much cheaper MP2 and adds the difference between CCSD(T) and MP2 calculated with the smaller basis set:
\begin{equation}
E^{\mathrm{focal}}_{\mathrm{CBS}} = E^{\infty}_{XY}[\mathrm{HF}] + E^{\mathrm{corr},\infty}_{XY}[\mathrm{MP2}] + E(\mathrm{CCSD(T)}/Y) - E(\mathrm{MP2}/Y),
\label{eq:schemeF}
\end{equation}
where $E^{\infty}_{XY}[\mathrm{HF}]$ and $E^{\mathrm{corr},\infty}_{XY}[\mathrm{MP2}]$ are the established two-point extrapolations of the HF energy and MP2 correlation energy, respectively\cite{Halkier1998_BasisSetConvergence,Halkier1999_BasisSetConvergence} (SI, p.~\pageref{si:schemes}). The last two terms are total energies evaluated with basis set $Y$. For comparison, we also apply the direct extrapolation of the CCSD(T) energy from the same two basis sets (Figure~\ref{fig:extrapolation-overview}b; SI, Eq.~\ref{eq:schemeD}).

We train one MLP on HF, MP2, CCSD, and CCSD(T) energies with the aug-cc-pVTZ and aug-cc-pVQZ basis sets and on HF, MP2, and CCSD forces with aug-cc-pVTZ, specifying the QM method through a descriptor as in our OMNI-P models\cite{AIO-OMNI-P1,OMNIP2x} (Figure~\ref{fig:extrapolation-overview}b). Progressively fewer geometries  supply training labels for the more expensive methods, so training does not require every QM method at every geometry. Neither extrapolated energies nor the schemes enter training; the reported predictions average five models. We assess the resulting extrapolated predictions against a theoretical best estimate (TBE) from true QM calculations with the larger aug-cc-pVQZ and aug-cc-pV5Z basis sets (SI, p.~\pageref{si:schemes}).

On the held-out test set, the focal-point MLP gives a mean absolute error of 0.017~kcal/mol in relative energies against the TBE, five times smaller than the 0.085~kcal/mol of the most accurate training QM method, CCSD(T)/aug-cc-pVQZ (Figure~\ref{fig:extrapolation-overview}c).
The same scheme applied to true QM energies gives 0.010~kcal/mol, showing that the MLP retains most of the accuracy gained by extrapolation. Thus, an MLP trained without extrapolated targets can exceed the accuracy of every method supplying its training data for the right reason (agreement for a barrier alone can be misleading: CCSD/aug-cc-pVQZ reproduces the ammonia barrier through cancellation of basis-set and correlation errors, despite larger errors over the surrounding PES, see SI, p.~\pageref{si:extrapolation-discussion}). This advantage persists along the relaxed inversion profile up to about 12~kcal/mol above the minimum, close to the edge of the training geometries (dotted lines in Figure~\ref{fig:extrapolation-overview}e).

The gain in accuracy also improves the tunneling splitting of ammonia, illustrating how extrapolation after training can benefit simulations at MLP cost: on the relaxed inversion profile, the focal-point MLP gives 0.97 times the splitting on the TBE surface, against 0.89 for the MLP at CCSD(T)/aug-cc-pVQZ (Figure~\ref{fig:extrapolation-overview}d). A single splitting requires about 5 million bead energy-and-force evaluations during instanton optimization, yet the full MLP calculation takes about 40 core-hours on Intel Xeon Gold 6240 CPUs, avoiding repeated QM calculations for the components of the focal-point scheme (SI, p.~\pageref{si:extrapolation-discussion}). In full-dimensional ring-polymer instanton calculations,\cite{Richardson2011_RingPolymerInstanton} the focal-point MLP gives a splitting of 0.754~cm$^{-1}$ and the MLP at CCSD(T)/aug-cc-pVQZ 0.669~cm$^{-1}$, against the experimental 0.793~cm$^{-1}$\cite{Urban1984_ForbiddenBandInversionRotation} (SI, Table~\ref{tab:si-nh3-instanton}). The approximations in the electronic surfaces and tunneling treatment are discussed in the SI (p.~\pageref{si:extrapolation-discussion}). Beyond ammonia, including full CI among the training methods allows the same construction to target the full CI/CBS limit for the \ce{H^{.} + H2} reaction (SI, p.~\pageref{si:h3-discussion}).

Because the extrapolation schemes are applied to the MLP's predictions only after training, the same MLP can supply predictions from different schemes without further QM calculations, and these schemes can differ in accuracy: on the held-out \ce{H^{.} + H2} geometries, direct extrapolation gives 0.022 and focal point 0.059~kcal/mol, while for ammonia the two are almost the same (0.018 and 0.017; SI, Tables~\ref{tab:si-h3-test} and~\ref{tab:si-nh3-test}). Their relative accuracy can be assessed against reference calculations chosen for the property of interest, e.g., at the minimum and saddle point for a barrier (SI, p.~\pageref{si:extrapolation-discussion}). Hence, the training dataset need not be constructed for one predetermined extrapolation scheme: a method-transferable MLP can reuse its learned QM methods in different schemes as the target property changes. However, when candidate schemes are known before training, we can readily include their prediction errors relative to the same schemes applied to the training QM energies in the loss, which brings the MLP to the accuracy of the same schemes on true QM energies (SI, p.~\pageref{si:scheme-term-discussion}).

\subsection{Putting QM methods on continuous coordinates in the method space}
\label{sec:coordinates}

MLPs are capable of generalizing in configurational space by adopting continuous coordinates: molecular descriptors as a function of nuclear coordinates.
We propose to extend this generalization to method space by giving MLPs continuous method descriptors informed by QM theory. The MLP can then learn relationships between QM methods and predict at methods absent from training. As a realization of our proposal, we can use parameters of common Kohn--Sham DFT functionals.

As a proof of concept, we use the fraction of exact (Hartree--Fock) exchange $a$ as a physically informed continuous method descriptor for a space of hybrid functionals. In particular, we use the functional family of Adamo and Barone,\cite{Adamo1997_TowardReliableAdiabatic} in which the remaining exchange is Becke's 1988 (B88) functional,\cite{Becke1988_DensityFunctionalExchangeEnergy} and the correlation is the Lee--Yang--Parr (LYP) functional:\cite{Lee1988_DevelopmentColleSalvettiCorrelation}
\begin{equation}
E_{xc}(a) = a\,E_x^{\mathrm{HF}} + (1-a)\,E_x^{\mathrm{B88}} + E_c^{\mathrm{LYP}}.
\label{eq:line}
\end{equation}

\begin{figure}[htbp]
\centering
\includegraphics[width=\linewidth]{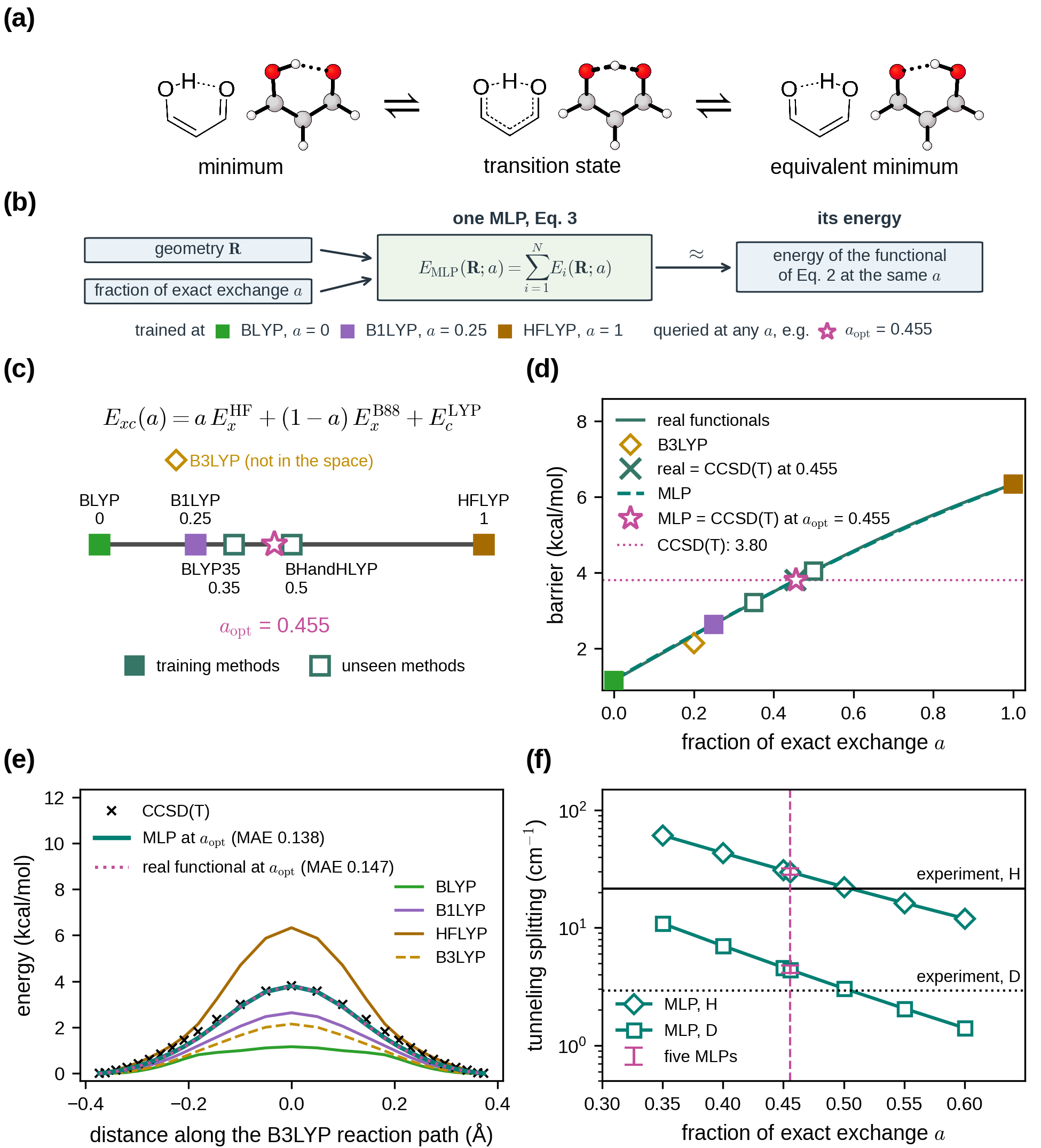}
\caption{\textbf{The fraction of exact exchange as the method descriptor}. (a) The proton transfer in malonaldehyde: the B3LYP minimum, transition state and equivalent minimum as 2D and 3D structures. (b) The MLP of Eq.~\ref{eq:mlp}, trained at three values of $a$ and queried at any $a$. (c) The space of Eq.~\ref{eq:line} with its five named functionals, B3LYP beside it, and $a_\mathrm{opt}$. (d) The barrier at the two B3LYP stationary geometries against $a$: the real functionals (DFT with Eq.~\ref{eq:line}), the MLP and CCSD(T), and where each meets CCSD(T). (e) The energy along the B3LYP reaction path, with the mean absolute error (MAE) against CCSD(T) over its 12 symmetry-unique geometries. (f) The tunneling splitting of malonaldehyde (H) and of its deuterated isotopologue (D) from ring-polymer instanton calculations on the MLP against $a$, the range over the ensemble's five MLPs at $a_\mathrm{opt}$, and experiment (H\cite{Baba1999_DetectionTunnelingRotation} and D\cite{Baughcum1984_MicrowaveSpectroscopicStudy}).~} %
\label{fig:coordinate}
\end{figure}

Several named functionals, including our recently proposed UMK26 ($a = 0.75$),\cite{Kumar2026_PrimaryRoleSemi} occupy points in this space (Figure~\ref{fig:coordinate}c). We use BLYP ($a = 0$), B1LYP ($a = 0.25$)\cite{Adamo1997_TowardReliableAdiabatic} and HFLYP ($a = 1$) for training, leaving BLYP35 ($a = 0.35$) and BHandHLYP ($a = 0.5$)\cite{Renz2009_ReliableQuantumChemical} as unseen functionals to test the predictions. The same construction can accommodate several parameters, as in B3LYP-type hybrids.\cite{Stephens1994_AbInitioCalculationVibrational}

As a demonstration, we construct an MLP for intramolecular proton transfer in malonaldehyde (Figure~\ref{fig:coordinate}a), whose barrier is sensitive to the fraction of exact exchange.  We repurpose the OMNI-P1 architecture\cite{AIO-OMNI-P1} by taking $a$ as the method descriptor alongside the molecular descriptor and train on energies and forces from the three functionals with the cc-pVDZ basis set. One MLP then predicts the energy as a function of both the geometry $\mathbf{R}$ and the QM method descriptor $a$ (Figure~\ref{fig:coordinate}b):
\begin{equation}
E_{\mathrm{MLP}}(\mathbf{R}; a) = \sum_{i=1}^{N} E_i(\mathbf{R}; a),
\label{eq:mlp}
\end{equation}
where $N$ is the number of atoms and $E_i$ is the atomic energy predicted from the chemical environment and $a$ by a network shared by atoms of the same element. Changing $a$ queries a different functional without retraining the MLP (SI, p.~\pageref{si:mlp}).

To choose a method more accurate than the training functionals, we find the optimal value $a_\mathrm{opt} = 0.455$ at which the MLP reproduces the CCSD(T)/cc-pVDZ reference barrier between the B3LYP/cc-pVDZ minimum and transition state (Figure~\ref{fig:coordinate}d). This requires two CCSD(T) single-point energies and no further DFT calculations. These energies select the method descriptor, while the geometry dependence of the surface is learned from the DFT training data. On relaxing the minimum and transition state with the selected MLP, the barrier error remains 0.05~kcal/mol, compared with 1.23~kcal/mol for the best training functional (SI, Table~\ref{tab:si-coordinate-relaxed}).

The optimal method descriptor $a_\mathrm{opt}$ defines an actual DFT functional, so we can check whether the MLP reproduces it. Along the B3LYP reaction path, the MLP differs from this functional by an average of 0.010~kcal/mol, and it also predicts the withheld BLYP35 and BHandHLYP functionals with errors comparable to those at the training functionals (SI, Table~\ref{tab:si-coordinate-fit}). The method descriptor thus retains its physical meaning at values not included in training. Away from the two geometries used to choose $a_\mathrm{opt}$, the MLP is also more accurate than every functional in its training set: its mean absolute error against CCSD(T) along the B3LYP reaction path is 0.14~kcal/mol, compared with 0.46--0.84~kcal/mol for the training functionals (Figure~\ref{fig:coordinate}e; SI, Table~\ref{tab:si-coordinate-path}).

The MLP is most useful in simulations that require many energy evaluations, such as tunneling splittings. Ring-polymer instanton calculations of one splitting require about 2.6 million bead energy-and-force evaluations, supplied by our MLP in about 13 core-hours on an Intel Xeon Gold 6240 CPU, compared with an estimated 25,000--32,000 core-hours using cc-pVDZ DFT. At $a_\mathrm{opt}$, our MLP trained on relatively inexpensive DFT data gives tunneling splittings of 29.8 and 4.39~cm$^{-1}$ for H and D (Figure~\ref{fig:coordinate}f), compared with the experimental values of 21.6 and 2.915~cm$^{-1}$, respectively.\cite{Baba1999_DetectionTunnelingRotation,Baughcum1984_MicrowaveSpectroscopicStudy} The same instanton theory applied to an MLP transfer-learned from MP2 to the more expensive CCSD(T) reference with the larger aug-cc-pVTZ basis set gives 25.3 and 3.7~cm$^{-1}$.\cite{Kaser2022_TransferLearningAffordable} Our method-transferable MLP thus gives splittings within 20\,\% of those of the dedicated single-method MLP.

We envision extending continuous method descriptors to wavefunction-based methods, so that an MLP can learn how electron correlation and basis-set completeness change together. The discrete correlation approximations and basis sets of the Pople diagram\cite{Pople1965_DimensionalChartQuantum} would need physically justified coordinates with a meaningful joint limit, such as full CI at the complete basis set limit. The MLP could then learn the relations between methods and predict at coordinates not included in training. We are working toward this extension by developing such coordinates and testing predictions at unseen methods and will report these developments separately.

\section{Outlook: toward universal method-transferable MLPs}
\label{sec:universal}
\label{sec:outlook}

Our demonstrations concern individual molecules, for which dense sampling of the configurational space allows the MLP to resolve accuracy gains beyond the training QM methods. Universal MLPs sample the potential energy surfaces of many compounds much more sparsely. Extending our constructions to them requires learning the relationships between methods across chemical compound space as well. Whether these relationships can be learned accurately enough to exceed the accuracy of the training QM methods remains an open question.

Recent universal MLPs make this question increasingly relevant. On selected ligand-strain and conformer benchmarks, UMA-M approaches direct DFT accuracy.\cite{Wood2025_UmaFamilyUniversal} For such applications, reducing the fitting error alone cannot systematically correct errors in the training QM method. Learning to predict at more accurate, unseen methods would give universal MLPs a way to address this limitation.

The constructions demonstrated here suggest how to pursue this goal. Training a universal MLP on the components of extrapolation schemes would allow the scheme to be chosen after training, as demonstrated for ammonia. Continuous method descriptors offer predictions between and beyond the training methods, with interpolation demonstrated here for malonaldehyde. Both constructions can use existing architectures with a method input or multiple outputs.

Method transfer can also improve universal MLPs when fitting errors remain substantial. An MLP aimed at a more accurate QM method can be more accurate than one aimed at a less accurate method even if neither reaches its target (Figure~\ref{fig:concept}b). Our universal MLP OMNI-P1 illustrates this within one model: changing its method descriptor from DFT to CCSD(T)*/CBS improves its accuracy, although both predictions remain less accurate than direct DFT calculations.\cite{AIO-OMNI-P1} This motivates pursuing more accurate, unseen methods even for inexpensive models with substantial fitting errors. The outputs could also correct HF or MP2 toward CCSD(T)*/CBS through $\Delta$-learning, extending the DFT-based construction demonstrated with OMNI-P1.\cite{AIO-OMNI-P1}

These aims should guide dataset construction and model evaluation. Labels from multiple QM methods, as in ANI-1ccx\cite{Smith2020_AniCcxAni} and MultiXC-QM9,\cite{Nandi2023_MultixcQmLarge} provide information about relationships between methods.\cite{QM-DB-review} Expensive high-level labels remain a constraint for method-transferable MLPs too, but complete method coverage at every geometry is unnecessary: cheaper calculations can cover many geometries, supplemented by high-level calculations at selected ones.\cite{Gardner2026_UnderstandingMultiFidelityTraining} Method descriptors informed by systematic relationships, such as those represented in Jacob's ladder and the Pople diagram, are another key research direction. Evaluation must distinguish reproduction of the training QM methods from accuracy against references more accurate than every training method, so that improvements beyond those methods can be established.

In this Perspective, we argued that MLPs can exceed the accuracy of their training QM methods by learning across method space and predicting at unseen methods. We demonstrated this for individual molecules by repurposing established extrapolation schemes and introducing continuous method descriptors. For universal MLPs, these routes could improve on single-method models even before exceeding the accuracy of their training QM methods. Learning across method space is poised to become as routine as learning across configurational and chemical compound spaces.

\section{Code and data availability}
\label{sec:code}
The machine learning potentials and simulations in this work were implemented in the open-source MLatom package, available under the Apache 2.0 license at \url{https://github.com/dralgroup/mlatom}. The paper-specific scripts, MLatom source needed to use the reported models, input files, trained models, training and test data, model predictions, and selected raw simulation logs and outputs are available together on Zenodo at \url{https://doi.org/10.5281/zenodo.23329249}. Original code is provided under the Apache 2.0 license, and original data, model parameters, and results under the CC BY 4.0 license; third-party components retain their original licenses.

\section{Authors contributions}
\label{sec:contributions}
P.O.D. conceived and directed the study, carried out the research and wrote the manuscript. Y.-F.H. proposed learning across the continuous space of DFT functionals and developed the active-learning protocol used in this work. Y.C. developed the extension of the OMNI-P architecture on which all MLPs in this work are based. H.N. and M.W. carried out a preliminary study of continuous descriptors of wavefunction-based methods, demonstrating the feasibility of constructing MLPs more accurate than the QM methods used to train them. S.\'S. and I.G. contributed through discussions on extrapolation schemes and on machine learning across the space of DFT functionals. P.O.D.
takes responsibility for all content. The calculations, analysis and drafting were carried out by the
AI Dral Group, the group's collective of specialised AI research agents, under the authors' direction
and verification, with every result reproducible from the repository by a single command.

\section{Declaration of generative AI use}
\label{sec:ai}
Generative AI was used substantively in this work. The study was executed by the AI Dral Group, an AI
research-agent collective that is independent of the underlying agentic runtime; in this work it
operated on the Claude Code agent with the Claude Opus~5.5 and Claude Fable~5.1 models (Anthropic), the Codex agent with the GPT-6 Astra and GPT-6 Sol models (OpenAI), and the Protomia agent (Aitomistic) with the Qwen3.8-27B model (Qwen team, Alibaba) run locally. The human authors
directed and verified every step and take full responsibility for the content, which is
machine-verifiable and reproducible from the committed repository.

\begin{acknowledgement}

P.O.D. acknowledges funding by the projects for International Senior Scientists (Project No.: W2531013) and for Outstanding Youth Scholars (Overseas, 2021) of the National Natural Science Foundation of China, via the Lab project of the State Key Laboratory of Physical Chemistry of Solid Surfaces. S.\'S. thanks the Polish National Science Center for the partial financial support under Grant No. 2021/42/E/ST4/00096. We acknowledge Polish high-performance computing infrastructure PLGrid for awarding this project access to the LUMI supercomputer, owned by the EuroHPC Joint Undertaking, hosted by CSC (Finland) and the LUMI consortium through PLL/2026/10/019484.

This work was supported by the PIAST-AI Factory project -- PIAST AIF, co-financed by the EuroHPC Joint Undertaking through the European Union's Digital Europe Programme, and by Poland as the Participating Country, under Grant Agreement No. 101250730. Views and opinions expressed are however those of the author(s) only and do not necessarily reflect those of the European Union, EuroHPC Joint Undertaking or Poland. Neither the European Union, the EuroHPC Join Undertaking nor Poland can be held responsible for them.

\end{acknowledgement}

\bibliography{beyond-training}
\startsireferences{beyond-training}

\clearpage

\setcounter{page}{1}\renewcommand{\thepage}{S\arabic{page}}
\setcounter{table}{0}\renewcommand{\thetable}{S\arabic{table}}
\setcounter{figure}{0}\renewcommand{\thefigure}{S\arabic{figure}}
\setcounter{equation}{0}\renewcommand{\theequation}{S\arabic{equation}}
\makeatletter
\let\si@table\table \renewcommand{\table}[1][]{\si@table[!htbp]}
\let\si@figure\figure \renewcommand{\figure}[1][]{\si@figure[!htbp]}
\makeatother
\renewcommand{\topfraction}{0.9}\renewcommand{\bottomfraction}{0.8}
\renewcommand{\textfraction}{0.07}\renewcommand{\floatpagefraction}{0.85}

\phantomsection\pdfbookmark[1]{Supporting Information}{si.titlepage}
\begin{center}
  {\Large Supporting Information for}\\[8pt]
  {\Large\bfseries Can Machine Learning Potentials Be More Accurate\\
   Than the Methods They Are Trained On?}\\[14pt]
  Pavlo O. Dral\textsuperscript{1,2,3,4,*}\\[3pt]
  Yi-Fan Hou,\textsuperscript{1} Yuxinxin Chen,\textsuperscript{1}
  Hassan Nawaz\textsuperscript{1} and Mateusz Witkowski\textsuperscript{2,3}\\[3pt]
  Szymon \'Smiga\textsuperscript{2,3} and Ireneusz Grabowski\textsuperscript{2,3}
\end{center}

\vspace{0.6em}
{\small\itshape
\noindent\textsuperscript{1}State Key Laboratory of Physical Chemistry of Solid Surfaces, College of Chemistry and Chemical Engineering, and Fujian Provincial Key Laboratory of Theoretical and Computational Chemistry, Xiamen University, Xiamen, Fujian 361005, China

\noindent\textsuperscript{2}Institute of Physics, Faculty of Physics, Astronomy, and Informatics, Nicolaus Copernicus University in Toru\'n, ul. Grudziadzka 5, 87-100 Toru\'n, Poland

\noindent\textsuperscript{3}Institute of Advanced Studies, Nicolaus Copernicus University in Toruń, ul. Wileńska 4, 87-100 Toruń, Poland

\noindent\textsuperscript{4}Aitomistic, Shenzhen 518000, China
\par}

\vspace{0.6em}
\noindent\textsuperscript{*}\,E-Mail: \href{mailto:dral@xmu.edu.cn}{dral@xmu.edu.cn}

\vspace{1.4em}
\printsicontents
\bigskip

\sisection{Computational details}
\label{si:methods}

\sisubsection{Programs and reference QM calculations}
\label{si:reference-calcs}

\paragraph{Programs.}
All calculations, unless stated otherwise, run through MLatom,\cite{MLatom3} in a development version that adds the MLP with the method descriptor and the active learning with several reference methods of this work, with Python~3.11, PyTorch~2.1.2 and TorchANI~2.2.3\cite{Gao2020_TorchaniFreeOpen} for the MLP, PySCF~2.4.0\cite{PySCF_SunRecentDevelopments} for DFT and CCSD(T), geomeTRIC~1.0.2\cite{Wang2016_GeometryOptimizationMadeSimple} for geometry optimizations, and i-PI~3.2.0\cite{Litman2024_IPiFlexible} for the ring-polymer instanton, with our own interface to the MLPs of MLatom and our own corrections to its tunneling-splitting post-processing. MLatom provides the IRC, the molecular dynamics (velocity Verlet), the Wigner and harmonic quantum Boltzmann sampling of initial conditions, the analytic Hessians of the MLP, and the averaging of the ensemble. In the extrapolation route, the QM labels are calculated with PySCF~2.4.0 directly, without MLatom.

\paragraph{Reference calculations of the extrapolation route.}
The QM calculations are done with PySCF~2.4\cite{PySCF_SunRecentDevelopments} and the cc-pVXZ\cite{Dunning1989_GaussianBasisSets} and aug-cc-pVXZ\cite{Kendall1992_ElectronAffinitiesFirstRow} basis sets, from the unrestricted HF solution for \ce{H^{.} + H2} and the restricted one for ammonia, with the N~1s orbital frozen in all correlated calculations. CCSD and CCSD(T) energies come from one CCSD solution per geometry and basis set, and full CI for \ce{H^{.} + H2} is a complete-active-space configuration interaction with all electrons in all orbitals. SCF energies are converged to $10^{-12}$ and CCSD energies to $10^{-10}$~Hartree. The analytic gradients of HF, MP2, and CCSD agree with central finite differences to $2 \times 10^{-6}$~Hartree/\AA\ for both systems, while the analytic CCSD(T) gradients of this version of PySCF deviate by $10^{-4}$ to $3 \times 10^{-3}$~Hartree/\AA, too much to be used as training labels; the MLP therefore learns the forces of HF, MP2, and CCSD and only the energies of CCSD(T) and full CI, and validation and test geometries are calculated with energies only. For ammonia, the aug-cc-pVQZ calculations, which are the most expensive of the training set, are done for half of the training geometries and with energies only, so that all its forces come from aug-cc-pVTZ. %

Every QM method is also calculated at one reference geometry: \ce{H2} at 0.7414~\AA, the equilibrium distance of the London--Eyring--Polanyi--Sato surface used to generate the geometries (p.~\pageref{si:training-data}), with the third atom 8~\AA\ away for \ce{H^{.} + H2}, and the pyramidal stationary point of the focal-point analysis\cite{Csaszar1998_PursuitAbInitioLimit} used to generate the geometries (N--H 1.012~\AA, H--N--H 106.7$^\circ$) for ammonia. The reference geometry is the same for every QM method and is not optimized with any of them, so some training geometries lie below it: on each surface, 27--53 of the \ce{H^{.} + H2} geometries, by at most 0.03~kcal/mol, and up to six of the ammonia geometries, by at most 0.26~kcal/mol with HF and 0.004~kcal/mol with the correlated methods. Every relative energy reported is an energy difference on one surface: the energy at a geometry minus the energy of the same surface at the reference geometry, for the MLP (its own prediction there, which is not exactly zero) as for every QM method.

\paragraph{Reference calculations of the interpolation route.}
DFT energies and forces in the space of Eq.~\ref{eq:line} are calculated with PySCF\cite{PySCF_SunRecentDevelopments} through MLatom,\cite{MLatom3} each functional given as a custom exchange--correlation functional, with the cc-pVDZ basis set\cite{Dunning1989_GaussianBasisSets} and PySCF's default integration grid and convergence thresholds. CCSD(T) energies are calculated with PySCF from the restricted Hartree--Fock reference, with the 1s orbitals of the carbon and oxygen atoms frozen; where the CCSD iterations stalled on a nearly singular DIIS subspace, they were restarted with a larger subspace and then without DIIS.  Relative energies in the comparisons are formed by subtracting the energy of the same surface at the B3LYP/cc-pVDZ minimum, one fixed geometry for every functional and for CCSD(T). The B3LYP/cc-pVDZ intrinsic reaction coordinate (IRC) used in the tests is calculated with MLatom from the B3LYP transition state in both directions, with the exact Hessian at every step and steps of 0.05~\AA; it has 27 points, the transition state and 13 on each side.

\FloatBarrier
\sisubsection{The method-transferable MLP}
\label{si:mlp}

\paragraph{Architecture and method descriptor.}
The MLP is an ANI-type neural network\cite{Smith2017_AniExtensibleNeural} with the atomic environment vectors of ANI-1x and atomic networks of 160, 128, and 96 neurons, as implemented in the TorchANI\cite{Gao2020_TorchaniFreeOpen} interface of MLatom,\cite{MLatom3} with the method descriptor appended to the atomic environment vector of every atom, as in our OMNI-P series of models.\cite{AIO-OMNI-P1,OMNIP2x} In the extrapolation route, the descriptor is a one-hot vector over the nine (\ce{H^{.} + H2}) or eight (ammonia) training QM methods (Table~\ref{tab:si-training-labels}). CCSD is among them although neither scheme uses it, because it comes out of the same coupled-cluster calculation as CCSD(T) at no extra cost.  Each training geometry contributes energy labels for its selected QM methods (Table~\ref{tab:si-training-labels}). Forces are also used as labels for HF, MP2 and CCSD, whose analytic gradients in PySCF~2.4.0 agree with finite differences of the energies (p.~\pageref{si:reference-calcs}). CCSD(T) and full CI are used with energies only: the analytic CCSD(T) gradients of this PySCF version deviate from finite differences by up to $3 \times 10^{-3}$~Hartree/\AA, and full CI has no analytic gradient in it. For ammonia, the aug-cc-pVQZ calculations, which take minutes per geometry, are done with energies only, so all forces of ammonia come from aug-cc-pVTZ. Each label is used together with the method descriptor of its QM method. In the interpolation route, all three training functionals are used with energies and forces.

In the interpolation route, the descriptor is the exact-exchange fraction $a$, and each atomic network returns one energy as in Eq.~\ref{eq:mlp}. Both routes use GELU activations,\cite{DanHendrycks2016_GaussianErrorLinear} whose smooth second derivatives are needed for frequencies and tunneling splittings.\cite{ANI-1ccx-gelu-anharmonic}

\begin{table}
\caption{Training QM methods, method descriptors and labels of the MLPs of both routes. $\mathbf{e}_k$ is the one-hot vector with 1 at position $k$ of 9 (\ce{H^{.} + H2}) or 8 (ammonia); the share is that of the training geometries  used for training with the QM method (percentages rounded to integers) (p.~\pageref{si:training-data} and p.~\pageref{si:coordinate-al}).\label{tab:si-training-labels}}
\small
\begin{tabular}{llccc}
\hline
system & QM method & descriptor & labels & share \\
\hline
\ce{H^{.} + H2} & HF/cc-pVTZ & $\mathbf{e}_1$ & energies, forces & 100\% \\
 & MP2/cc-pVTZ & $\mathbf{e}_2$ & energies, forces & 100\% \\
 & CCSD/cc-pVTZ & $\mathbf{e}_3$ & energies, forces & 60\% \\
 & CCSD(T)/cc-pVTZ & $\mathbf{e}_4$ & energies & 60\% \\
 & full CI/cc-pVTZ & $\mathbf{e}_5$ & energies & 25\% \\
 & HF/cc-pVQZ & $\mathbf{e}_6$ & energies, forces & 63\% \\
 & MP2/cc-pVQZ & $\mathbf{e}_7$ & energies, forces & 63\% \\
 & CCSD/cc-pVQZ & $\mathbf{e}_8$ & energies, forces & 38\% \\
 & CCSD(T)/cc-pVQZ & $\mathbf{e}_9$ & energies & 38\% \\
\hline
ammonia & HF/aug-cc-pVTZ & $\mathbf{e}_1$ & energies, forces & 100\% \\
 & MP2/aug-cc-pVTZ & $\mathbf{e}_2$ & energies, forces & 100\% \\
 & CCSD/aug-cc-pVTZ & $\mathbf{e}_3$ & energies, forces & 60\% \\
 & CCSD(T)/aug-cc-pVTZ & $\mathbf{e}_4$ & energies & 60\% \\
 & HF/aug-cc-pVQZ & $\mathbf{e}_5$ & energies & 50\% \\
 & MP2/aug-cc-pVQZ & $\mathbf{e}_6$ & energies & 50\% \\
 & CCSD/aug-cc-pVQZ & $\mathbf{e}_7$ & energies & 30\% \\
 & CCSD(T)/aug-cc-pVQZ & $\mathbf{e}_8$ & energies & 30\% \\
\hline
malonaldehyde & BLYP/cc-pVDZ & $a = 0$ & energies, forces & 100\% \\
 & B1LYP/cc-pVDZ & $a = 0.25$ & energies, forces & 100\% \\
 & HFLYP/cc-pVDZ & $a = 1$ & energies, forces & 100\% \\
\hline
\end{tabular}
\end{table}

\paragraph{Training.}
The MLP is trained with the defaults of MLatom's TorchANI interface: force weight 0.1, and  AdamW for network weights and SGD for biases, both with an initial learning rate of $10^{-3}$ halved after 64 epochs without improvement of the validation error until  the AdamW learning rate falls below $10^{-5}$.  In the extrapolation route, we use batches of 128 instead of 8 geometries and disable per-batch logging. The ammonia MLP of the main text is trained on the energies and forces of the training QM methods only.

In the interpolation route, the MLP is trained on energies and forces with the settings of our active-learning protocol:\cite{AL} the defaults of MLatom's TorchANI interface above, including batches of 8 geometries, with 10\% of the training set for validation instead of the interface's 20\%.

\paragraph{Energy zero and atomic energies.}
\label{si:energy-zero}
In both routes, the MLP gives the energy of QM method $m$, with method descriptor $d_m$ ($\mathbf{e}_m$ or $a$), at the geometry $\mathbf{R}$ as
\begin{equation}
E^\mathrm{MLP}_m(\mathbf{R}) = \sum_i E_i(\mathbf{R}, d_m) + \sum_i \varepsilon_{Z_i},
\label{eq:si-energy-zero}
\end{equation}
where $E_i$ is the output of the atomic network of atom $i$ and $\varepsilon_Z$ is one atomic energy per element, the same for all QM methods. The two routes differ only in the energy label $y_m(\mathbf{R})$ that Eq.~\ref{eq:si-energy-zero} is trained to reproduce:
\begin{equation}
y_m(\mathbf{R}) = E_m(\mathbf{R}) - E_m(\mathbf{R}_\mathrm{ref}) \quad \text{(extrapolation)}, \qquad y_m(\mathbf{R}) = E_m(\mathbf{R}) \quad \text{(interpolation)},
\end{equation}
with $E_m$ the energy of QM method $m$ and $\mathbf{R}_\mathrm{ref}$ the reference geometry (p.~\pageref{si:reference-calcs}). Before training, MLatom's TorchANI interface fits the $\varepsilon_Z$ by least squares to all labels of all QM methods at once, so that the networks learn only the much smaller remainder: they are trained on $y_m(\mathbf{R}) - \sum_i \varepsilon_{Z_i}$; in prediction, $\sum_i \varepsilon_{Z_i}$ is added back as in Eq.~\ref{eq:si-energy-zero}, and nothing else is: the extrapolation MLP predicts energies relative to the reference geometry and the interpolation MLP total energies. Since all geometries of a route are the same molecule, $\sum_i \varepsilon_{Z_i}$ is one constant, and every energy reported is a difference between two geometries on one surface, from which this constant and $E_m(\mathbf{R}_\mathrm{ref})$ both cancel. The schemes act on these differences unchanged because the coefficients with which each scheme combines the energies of the QM methods (Eq.~\ref{eq:increments}) sum to one. The subtraction of $E_m(\mathbf{R}_\mathrm{ref})$ in the extrapolation route was an early design choice, which, in principle, is not required.

\paragraph{Angular cutoff and ensemble.}
In the extrapolation route, the angular cutoff of the atomic environment vectors is chosen by five-fold cross-validation among 3.3, 3.5, and 4.0~\AA\ with the four angular shifts spread evenly from 0.9~\AA\ to the cutoff: the five folds are fixed once for all training geometries, and the five MLPs of the chosen cutoff, each validated on a different fold, form the ensemble whose averaged predictions are reported. The mean validation errors of the three cutoffs are 0.026--0.030~kcal/mol for the \ce{H^{.} + H2} reaction and 0.006--0.007~kcal/mol for ammonia (Tables~\ref{tab:si-h3-cutoff} and~\ref{tab:si-nh3-cutoff}). These differences are smaller than the standard deviations of the errors over the folds, 0.003--0.011 and 0.001--0.002~kcal/mol, so the choice among them does not change the validation error measurably. We use the cutoff with the lowest mean error, 3.5~\AA\ for the \ce{H^{.} + H2} reaction and 3.3~\AA\ for ammonia.
In the interpolation route, the angular cutoff of the atomic environment vectors is 4.0~\AA, with the four angular shifts spread evenly from 0.9~\AA\ to the cutoff. It was chosen once, by five-fold cross-validation among 3.3, 3.5 and 4.0~\AA\ on 450 geometries of an early design study of the initial data set, and was not chosen again for the reported data set (Table~\ref{tab:si-coordinate-cutoff}). The three cutoffs give energy errors of 0.165--0.192~kcal/mol, differences smaller than the standard deviations over the folds, 0.030--0.055~kcal/mol; 4.0~\AA, which has the lowest error, is also the value of our transferable models.\cite{AIQM1,AIQM2}

\begin{table}[h]
\caption{Choice of the angular cutoff of the MLP for the \ce{H^{.} + H2} reaction by five-fold cross-validation: MAE of relative energies on the validation fold in kcal/mol, averaged over the folds, for each training QM method and their mean (standard deviation over the folds in parentheses). The five MLPs of the chosen cutoff (3.5~\AA) form the ensemble.}
\label{tab:si-h3-cutoff}
\small
\begin{tabular}{lrrr}
\hline
training QM method & 3.3~\AA & 3.5~\AA & 4.0~\AA \\
\hline
HF/cc-pVTZ & 0.034 & 0.026 & 0.031 \\
MP2/cc-pVTZ & 0.022 & 0.022 & 0.023 \\
CCSD/cc-pVTZ & 0.024 & 0.022 & 0.027 \\
CCSD(T)/cc-pVTZ & 0.028 & 0.029 & 0.031 \\
full CI/cc-pVTZ & 0.033 & 0.031 & 0.040 \\
HF/cc-pVQZ & 0.027 & 0.029 & 0.026 \\
MP2/cc-pVQZ & 0.024 & 0.021 & 0.025 \\
CCSD/cc-pVQZ & 0.032 & 0.032 & 0.034 \\
CCSD(T)/cc-pVQZ & 0.030 & 0.028 & 0.032 \\
\hline
mean & 0.028 (0.011) & 0.026 (0.009) & 0.030 (0.003) \\
folds & 5 & 5 & 5 \\
\hline
\end{tabular}
\end{table}
\begin{table}[h]
\caption{Choice of the angular cutoff of the MLP for the ammonia by five-fold cross-validation: MAE of relative energies on the validation fold in kcal/mol, averaged over the folds, for each training QM method and their mean (standard deviation over the folds in parentheses). The five MLPs of the chosen cutoff (3.3~\AA) form the ensemble.}
\label{tab:si-nh3-cutoff}
\small
\begin{tabular}{lrrr}
\hline
training QM method & 3.3~\AA & 3.5~\AA & 4.0~\AA \\
\hline
HF/aug-cc-pVTZ & 0.005 & 0.005 & 0.006 \\
MP2/aug-cc-pVTZ & 0.004 & 0.005 & 0.005 \\
CCSD/aug-cc-pVTZ & 0.005 & 0.005 & 0.004 \\
CCSD(T)/aug-cc-pVTZ & 0.008 & 0.009 & 0.008 \\
HF/aug-cc-pVQZ & 0.009 & 0.012 & 0.010 \\
MP2/aug-cc-pVQZ & 0.006 & 0.007 & 0.008 \\
CCSD/aug-cc-pVQZ & 0.007 & 0.008 & 0.006 \\
CCSD(T)/aug-cc-pVQZ & 0.007 & 0.007 & 0.006 \\
\hline
mean & 0.006 (0.001) & 0.007 (0.002) & 0.007 (0.001) \\
folds & 5 & 5 & 5 \\
\hline
\end{tabular}
\end{table}
\begin{table}[h]
\caption{Choice of the angular cutoff of the atomic environment vectors of the malonaldehyde MLP: root-mean-square error (RMSE) of the energies in five-fold cross-validation and its standard deviation over the five folds, in kcal/mol, on 450 geometries of an early design study of the initial data set, labelled with the three training functionals. These geometries are not the training set of the reported MLP. The four angular shifts are spread evenly from 0.9~\AA\ to the cutoff.}
\label{tab:si-coordinate-cutoff}
\begin{tabular}{ccc}
\hline
cutoff (\AA) & RMSE & standard deviation over the folds \\
\hline
3.3 & 0.179 & 0.030 \\
3.5 & 0.192 & 0.038 \\
4.0 & 0.165 & 0.055 \\
\hline
\end{tabular}
\end{table}

\FloatBarrier
\sisubsection{Active learning}
\label{si:al}
Both routes select training geometries by our physics-informed active learning\cite{AL} in MLatom.\cite{MLatom3} In each iteration, the MLP runs a set of simulations; each molecular-dynamics trajectory is stopped at the first geometry at which the MLP is uncertain, and only this geometry is calculated with QM and added to the training set. A geometry optimization likewise contributes at most the first uncertain geometry along its path. The MLP is then retrained and the next iteration starts.
The uncertainty is the difference between the energies of a main MLP, trained on energies and forces, and an auxiliary MLP trained on energies only; a geometry is uncertain when the difference exceeds a threshold, by default the median plus three median absolute deviations of that difference over the validation set. The threshold, the simulations, the initial data set and the end of the loop are specific to each system (pp.~\pageref{si:training-data} and~\pageref{si:coordinate-al}).

\FloatBarrier
\sisubsection{Tunneling splittings by ring-polymer instanton theory}
\label{si:instanton-methods}

\paragraph{Frequencies and tunneling splittings.}
Except for the rate constants of the \ce{H^{.} + H2} reaction (p.~\pageref{si:rates-methods}), harmonic frequencies are obtained from the analytic Hessian of the ensemble, and zero-point energies as half the sum of the real frequencies. Tunneling splittings of malonaldehyde and ammonia are calculated with ring-polymer instanton theory\cite{Richardson2011_RingPolymerInstanton} in i-PI~3.2,\cite{Litman2024_IPiFlexible} following the protocol of K{\"a}ser, Richardson and Meuwly for malonaldehyde, with the same scripts for both molecules.\cite{Kaser2022_TransferLearningAffordable} i-PI calls the MLP directly in its own process, through its Python force-field interface, without a socket or an external driver. The reactant is the harmonic analysis of one minimum of the MLP from its analytic Hessian, with the masses and conventions of i-PI (i-PI's own harmonic analysis by finite differences of the forces puts the splitting 0.6\% higher), and the splitting is twice the tunneling matrix element of the instanton in the splitting mode of i-PI.

The half ring polymer between the two minima of the MLP is optimized with the L-BFGS algorithm and i-PI's convergence thresholds tightened tenfold, first with 64 beads at 100~K from the straight line between the minima, then with 256 beads at 50~K, and finally with 1024 beads at 25~K, each stage starting from the previous instanton interpolated to the new number of beads. The atoms carry the masses of their most abundant isotopes, except where an isotopologue is named.
The two molecules differ in the surface, in the optimization of the minima and in the Hessians of the beads, which are given for malonaldehyde on p.~\pageref{si:coordinate-discussion} and for ammonia below.

\paragraph{Fluctuation factor.}
The fluctuation factor needs the determinant of the Hessian of the whole half ring polymer, a matrix of $3n \times 1024$ rows for a molecule of $n$ atoms; we compute it by a block LU factorization of this block-tridiagonal matrix instead of i-PI's dense diagonalization, which on three test runs agrees with a dense calculation to $3 \times 10^{-10}$ in the logarithm of the determinant (Table~\ref{tab:si-coordinate-splitting-convergence}).

\paragraph{Inversion splittings of ammonia.}
\label{si:nh3-1d}
On each surface, in true QM and from the MLP alike, the relaxed inversion path consists of 34 values of the H--N--H angle from 78$^\circ$ to the planar structure (120$^\circ$), 2$^\circ$ apart below 96$^\circ$ and 1$^\circ$ above, with the N--H distance relaxed at each by a parabola through three points; the path is parametrized by the mass-weighted arc length with the motion of the centre of mass removed and mirrored at the planar structure, and the one-dimensional Schr{\"o}dinger equation of this symmetric double well is solved by finite differences on 1201 grid points; the splitting is the difference between the two lowest levels, and it changes by less than 0.1\% when the path is shortened by 4\% at each end. The same treatment is applied to every surface. It overestimates the experimental splitting of \ce{NH3} (0.79~cm$^{-1}$) by a factor of 1.8 on the TBE surface, so the surfaces are compared through the ratio of their splittings to that of the TBE surface.

The full-dimensional splittings are calculated with ring-polymer instanton theory\cite{Richardson2011_RingPolymerInstanton} with the same driver as for malonaldehyde (i-PI, above): a half ring polymer of 1024 beads at 25~K, reached through 64 beads at 100~K and 256 at 50~K, the ensemble evaluated in double precision, the bead Hessians analytic, the zero mode removed by projecting out the tangent of the path, the minimum optimized by Newton steps with the analytic Hessian of the surface, and the isotope masses of the one-dimensional treatment.
Doubling the number of beads to 2048 changes the splittings by $-0.4$\,\% (\ce{NH3}) and $-0.2$\,\% (\ce{ND3}), and halving the temperature to 12.5~K at the same bead spacing (2048 beads) changes them by $+0.4$\,\% and $+0.3$\,\%; the 1024-bead values are thus within 0.6\,\% of the bead limit.

\FloatBarrier
\sisubsection{Rate constants}
\label{si:rates-methods}
The \ce{H^{.} + H2 -> H2 + H^{.}} reaction is degenerate and has no measurable rate constant; the rate constants are therefore those of the \ce{D^{.} + H2 -> HD + H^{.}} reaction, which runs on the same potential energy surface with different nuclear masses and for which experiment and accurate quantum dynamics on the basis-set-limit surface of the same reaction\cite{Mielke2002_HierarchicalFamilyGlobal} agree over 167--2112~K.\cite{Mielke2003_HH2ThermalReaction}
On every surface, the collinear symmetric saddle point and its Hessian are obtained from the energies at the same 22 geometries. The surfaces are each training QM method and each scheme, in true QM and from the MLP (the averaged energies of the ensemble) alike, and the TBE in true QM.

\begin{figure}[htbp]
\centering
\includegraphics[width=\linewidth]{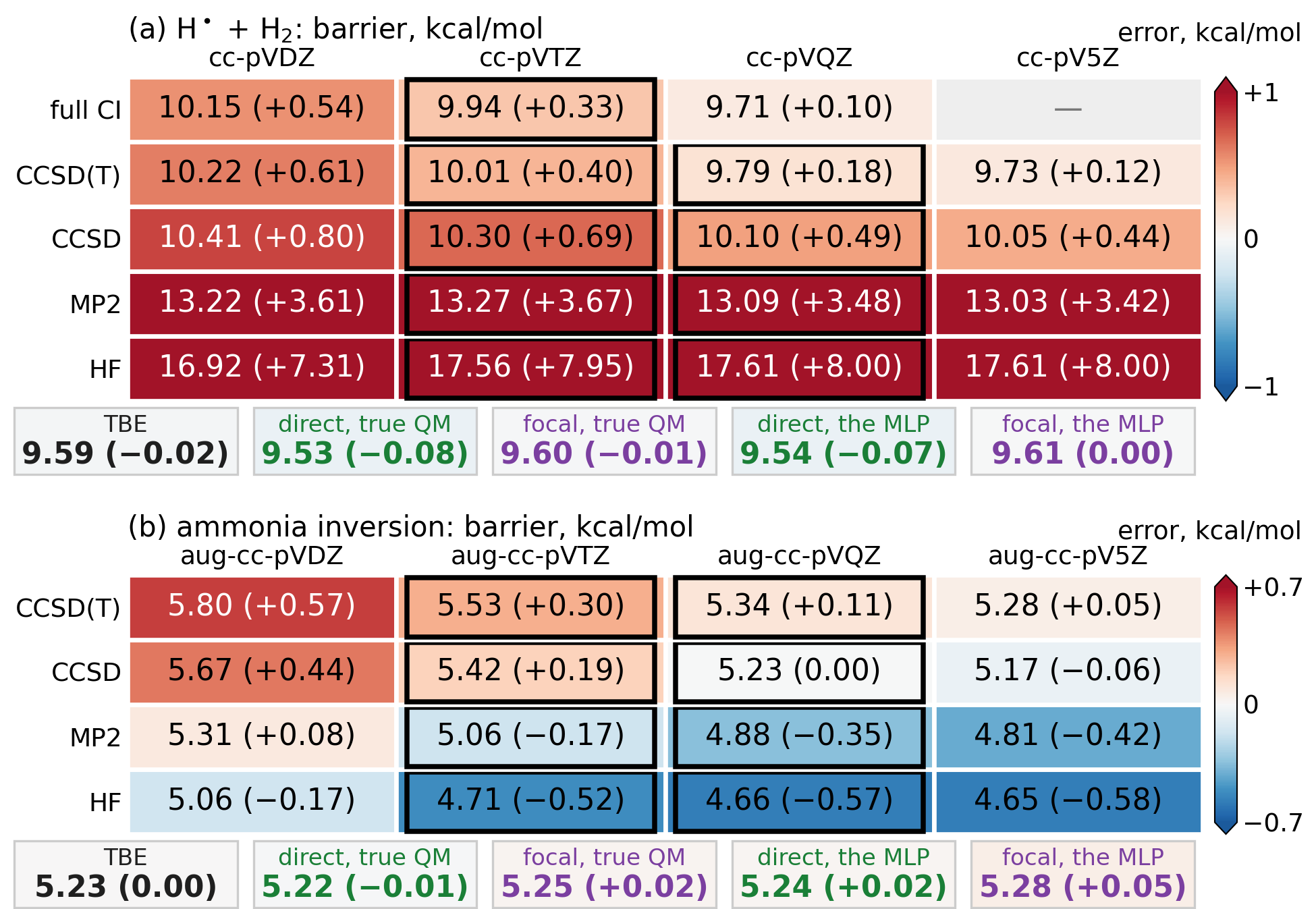}
\caption{\textbf{Two extrapolation schemes applied to the same training QM methods, in true QM and on the MLP.} (a) The barrier of the \ce{H^{.} + H2} reaction in kcal/mol with every QM method and the cc-pVXZ basis sets at the benchmark saddle geometry ($r_{\mathrm{HH}} = 0.930$~\AA),\cite{Anderson2016_ExactBarrierHH2} in parentheses and in color the error against the exact barrier of 9.6089~kcal/mol; the training QM methods are framed (full CI with cc-pV5Z was not calculated). Below the table: the TBE and both schemes applied to the true QM energies and to the MLP's, all including $\Delta E_{\mathrm{FCI}}$. (b) The same for the inversion barrier of ammonia with the augmented basis sets at the geometries of the original focal-point analysis,\cite{Csaszar1998_PursuitAbInitioLimit} with the errors against the TBE.}
\label{fig:extrapolation}
\end{figure}

The 22 geometries are five points along the symmetric stretch (H--H distances from 0.91 to 0.95~\AA), the middle atom displaced along the axis and perpendicular to it at three of those five, and \ce{H2} at five bond lengths with the third atom 8~\AA\ away. The saddle point is the minimum of a quartic polynomial fitted along the symmetric stretch, and the three curvatures that fix the Cartesian Hessian of the linear symmetric \ce{H-H-H} are obtained from the displaced energies by Richardson-extrapolated finite differences and interpolated to the saddle point. The barriers quoted in the text are those between each surface's own stationary points, which differ by up to  0.05~kcal/mol from the barriers at the fixed benchmark geometry of SI, Figure~\ref{fig:extrapolation} (SI, Tables~\ref{tab:si-h3-saddle} and~\ref{tab:si-nh3-splitting}).

The harmonic frequencies of \ce{D-H-H} at the saddle point and of \ce{H2} are obtained with the harmonic analysis of PySCF and the isotope masses. The rate constants of \ce{D^{.} + H2 -> HD + H^{.}} are calculated with conventional transition-state theory\cite{Eyring1935_ActivatedComplexChemical} with rigid-rotor and harmonic-oscillator partition functions and the tunneling correction of a symmetric Eckart barrier\cite{Eckart1930_PenetrationPotentialBarrier} with the zero-point-corrected barrier height and the imaginary frequency of the saddle point. The absolute rates on a model surface with the exact barrier differ from the published accurate quantum-dynamics rates\cite{Fleming2011_KineticIsotopeEffects} by factors of 1.2, 1.4, 2.8, and 13 at 1000, 500, 300, and 200~K.
Those published rates use a Born--Huang surface, whereas our calculations omit the diagonal Born--Oppenheimer correction, which raises the barrier of this reaction by 0.14~kcal/mol;\cite{Mielke2005_BornOppenheimerDiagonalCorrection} the factors therefore mix differences in the electronic surface with those in the kinetic treatment. We compare surfaces through rates calculated under the same treatment, relative to the TBE surface.

How much of the treatment cancels in such a ratio is itself measurable: replacing the Eckart correction by the Wigner one changes the ratios quoted here by 0.2\% at 1000~K, 1.2\% at 500~K, 5.5\% at 300~K and 17\% at 200~K, so the cancellation is good where tunneling is a correction and only partial at the lowest temperature, where it is not.

\FloatBarrier
\clearpage
\sisection{Extrapolation after training}
\label{si:methods-extrapolation}
\label{si:extrapolation-mlp}

\sisubsection{The schemes and the TBE}
\label{si:schemes}

\paragraph{The schemes.}
The direct-extrapolation scheme extrapolates the HF energy and the CCSD(T) correlation energy:
\begin{equation}
E^{\mathrm{direct}}_{\mathrm{CBS}} = E^{\infty}_{XY}[\mathrm{HF}] + E^{\mathrm{corr},\infty}_{XY}[\mathrm{CCSD(T)}].
\label{eq:schemeD}
\end{equation}
In both extrapolation schemes (Eqs.~\ref{eq:schemeF} and~\ref{eq:schemeD}), the correlation energy of a QM method $M$ calculated with two basis sets with the cardinal numbers $X > Y$ is extrapolated with the two-point formula of Halkier et al.:\cite{Halkier1998_BasisSetConvergence}
\begin{equation}
E^{\mathrm{corr},\infty}_{XY}[M] = \frac{X^3 E^{\mathrm{corr}}_X[M] - Y^3 E^{\mathrm{corr}}_Y[M]}{X^3 - Y^3},
\label{eq:twopoint}
\end{equation}
where $E^{\mathrm{corr}}_X[M] = E(M/X) - E(\mathrm{HF}/X)$, and the HF energy with the two-point formula developed for it:\cite{Halkier1999_BasisSetConvergence}
\begin{equation}
E^{\infty}_{XY}[\mathrm{HF}] = \frac{e^{-\alpha Y} E(\mathrm{HF}/X) - e^{-\alpha X} E(\mathrm{HF}/Y)}{e^{-\alpha Y} - e^{-\alpha X}},
\label{eq:hfcbs}
\end{equation}
with $\alpha = 1.63$. For $X = 4$ and $Y = 3$, both schemes reduce to the same linear combination,
\begin{equation}
E_{\mathrm{CBS}} \approx E(\mathrm{CCSD(T)}/Y) \; \left[ + \Delta E_{\mathrm{FCI}} \right] + 1.73 \, \Delta_{YX}[M] - 0.49 \, \Delta_{YX}[\mathrm{HF}],
\label{eq:increments}
\end{equation}
where the optional $\Delta E_{\mathrm{FCI}} = E(\mathrm{full~CI}/Y) - E(\mathrm{CCSD(T)}/Y)$ is added to both schemes for the \ce{H^{.} + H2} reaction (p.~\pageref{si:h3-discussion}), for which full CI is affordable, so that both target full CI/CBS; $\Delta_{YX}[M] = E(M/X) - E(M/Y)$ is the change of the energy of a QM method $M$ from the smaller to the larger basis set, and $M$ is CCSD(T) in the direct-extrapolation scheme and MP2 in the focal-point scheme; the coefficients are rounded from $64/37 = 1.7297$ and $64/37 - e^{\alpha}/(e^{\alpha} - 1) = 0.4861$.
Only these changes are amplified: an error of the MLP that is the same for the two basis sets of a QM method cancels in them (SI, Tables~\ref{tab:si-h3-sharing} and~\ref{tab:si-nh3-sharing}), and what remains is the error of the first term, which enters once.

\begin{table}[h]
\caption{Errors of the MLP (the ensemble's averaged prediction) against true QM on the validation set of the \ce{H^{.} + H2} reaction for pairs of training QM methods: the MAE of the first and of the second QM method, the MAE of their difference, and the correlation coefficient of the two errors; MAEs in kcal/mol. A difference whose error is smaller than that of either QM method means an error shared by both, which the schemes do not amplify.}
\label{tab:si-h3-sharing}
\small
\begin{tabular}{lrrrr}
\hline
pair of QM methods & first & second & difference & correlation \\
\hline
full CI/cc-pVTZ $-$ CCSD(T)/cc-pVTZ & 0.010 & 0.010 & 0.006 & 0.95 \\
CCSD(T)/cc-pVTZ $-$ CCSD/cc-pVTZ & 0.010 & 0.014 & 0.015 & 0.67 \\
CCSD(T)/cc-pVQZ $-$ CCSD/cc-pVQZ & 0.011 & 0.013 & 0.013 & 0.56 \\
MP2/cc-pVQZ $-$ MP2/cc-pVTZ & 0.012 & 0.010 & 0.005 & 0.95 \\
CCSD(T)/cc-pVQZ $-$ CCSD(T)/cc-pVTZ & 0.011 & 0.010 & 0.004 & 0.98 \\
HF/cc-pVQZ $-$ HF/cc-pVTZ & 0.014 & 0.010 & 0.008 & 0.88 \\
\hline
\end{tabular}
\end{table}
\begin{table}[h]
\caption{Errors of the MLP (the ensemble's averaged prediction) against true QM on the validation set of the ammonia inversion for pairs of training QM methods: the MAE of the first and of the second QM method, the MAE of their difference, and the correlation coefficient of the two errors; MAEs in kcal/mol. A difference whose error is smaller than that of either QM method means an error shared by both, which the schemes do not amplify.}
\label{tab:si-nh3-sharing}
\small
\begin{tabular}{lrrrr}
\hline
pair of QM methods & first & second & difference & correlation \\
\hline
CCSD(T)/aug-cc-pVTZ $-$ CCSD/aug-cc-pVTZ & 0.008 & 0.003 & 0.007 & 0.61 \\
CCSD(T)/aug-cc-pVQZ $-$ CCSD/aug-cc-pVQZ & 0.009 & 0.007 & 0.002 & 0.97 \\
MP2/aug-cc-pVQZ $-$ MP2/aug-cc-pVTZ & 0.004 & 0.002 & 0.004 & 0.50 \\
CCSD(T)/aug-cc-pVQZ $-$ CCSD(T)/aug-cc-pVTZ & 0.009 & 0.008 & 0.004 & 0.88 \\
HF/aug-cc-pVQZ $-$ HF/aug-cc-pVTZ & 0.008 & 0.003 & 0.008 & 0.34 \\
\hline
\end{tabular}
\end{table}

\paragraph{The TBE.}
The TBE is the direct-extrapolation scheme with the next larger pair of basis sets ($X = 5$, $Y = 4$), so it shares its rule with one of the two schemes it scores; its focal-point form differs from it by 0.012~kcal/mol on average over the validation geometries of the \ce{H^{.} + H2} reaction, enough to unsettle the ordering of the two schemes but not their distance from the training QM methods. For the \ce{H^{.} + H2} reaction, $\Delta E_{\mathrm{FCI}}$ enters the schemes and the TBE alike from cc-pVTZ; with cc-pVQZ instead, its contribution to the barrier differs by 0.001~kcal/mol.

\paragraph{The schemes applied to true QM energies.}
\label{si:extrapolation}
Tables~\ref{tab:si-h3-ladder} and~\ref{tab:si-nh3-ladder} give the barrier with every QM method and basis set, and Table~\ref{tab:si-schemes} the barriers from the two schemes and the TBE. All values in these tables are calculated with true QM methods (PySCF; frozen core for ammonia).

\begin{table}[h]
\caption{Barrier of the \ce{H^{.} + H2 -> H2 + H^{.}} reaction in kcal/mol calculated with true QM methods at one saddle-point geometry ($r_{\mathrm{HH}} = 0.930$~\AA) relative to \ce{H^{.} + H2} ($r_{\mathrm{HH}} = 0.7414$~\AA); unrestricted HF reference for the open-shell species. The exact barrier is 9.61~kcal/mol. Full CI is exact for a given basis set, i.e., its error is only due to the incompleteness of the basis set. Dashes: not calculated.}
\label{tab:si-h3-ladder}
\small
\begin{tabular}{lrrrrrrrr}
\hline
 & \multicolumn{4}{c}{cc-pVXZ} & \multicolumn{4}{c}{aug-cc-pVXZ} \\
QM method & DZ & TZ & QZ & 5Z & DZ & TZ & QZ & 5Z \\
\hline
HF & 16.92 & 17.56 & 17.61 & 17.61 & 16.81 & 17.57 & 17.61 & 17.61 \\
MP2 & 13.22 & 13.27 & 13.09 & 13.03 & 12.56 & 13.10 & 13.01 & 12.99 \\
CCSD & 10.41 & 10.30 & 10.10 & 10.05 & 9.49 & 10.07 & 10.02 & 10.02 \\
CCSD(T) & 10.22 & 10.01 & 9.79 & 9.73 & 9.26 & 9.77 & 9.70 & 9.70 \\
full CI & 10.15 & 9.94 & 9.71 & -- & 9.18 & 9.69 & 9.63 & -- \\
\hline
\end{tabular}
\end{table}

\begin{table}[h]
\caption{Inversion barrier of ammonia in kcal/mol calculated with true QM methods (frozen core) at the reference geometries of the original focal-point analysis.}
\label{tab:si-nh3-ladder}
\small
\begin{tabular}{lrrrrrrrr}
\hline
 & \multicolumn{4}{c}{cc-pVXZ} & \multicolumn{4}{c}{aug-cc-pVXZ} \\
QM method & DZ & TZ & QZ & 5Z & DZ & TZ & QZ & 5Z \\
\hline
HF & 7.16 & 5.30 & 4.97 & 4.72 & 5.06 & 4.71 & 4.66 & 4.65 \\
MP2 & 7.98 & 5.92 & 5.37 & 4.96 & 5.31 & 5.06 & 4.88 & 4.81 \\
CCSD & 8.27 & 6.18 & 5.63 & 5.29 & 5.67 & 5.42 & 5.23 & 5.17 \\
CCSD(T) & 8.51 & 6.36 & 5.78 & 5.41 & 5.80 & 5.53 & 5.34 & 5.28 \\
\hline
\end{tabular}
\end{table}

\begin{table}[h]
\caption{Barriers from the two extrapolation schemes (Eqs.~\ref{eq:schemeD} and~\ref{eq:schemeF}) applied to the true QM energies with the TZ and QZ basis sets, and the theoretical best estimate (TBE), Eq.~\ref{eq:schemeD} with the QZ and 5Z basis sets, all in kcal/mol. For the \ce{H^{.} + H2} reaction, all three include the difference between full CI and CCSD(T) calculated with the TZ basis set (Eq.~\ref{eq:increments}), so that they approximate full CI/CBS; with the QZ basis set, this difference changes the barrier by 0.001~kcal/mol. The exact barrier of this reaction is 9.6089~kcal/mol: the TBE is 0.02~kcal/mol from it with cc-pVXZ, and the focal-point scheme happens to be closer at this one geometry (0.01~kcal/mol). The TBE is the reference of the SI because it can be calculated at every geometry; the exact value is known only for the barrier. Ammonia: at the reference geometries of Table~\ref{tab:si-nh3-ladder}.}
\label{tab:si-schemes}
\small
\begin{tabular}{lrrrr}
\hline
 & \multicolumn{2}{c}{\ce{H^{.} + H2} (with full CI $-$ CCSD(T))} & \multicolumn{2}{c}{ammonia} \\
 & cc-pVXZ & aug-cc-pVXZ & cc-pVXZ & aug-cc-pVXZ \\
\hline
direct extrapolation & 9.53 & 9.56 & 5.52 & 5.22 \\
focal point & 9.60 & 9.52 & 5.56 & 5.25 \\
TBE & 9.59 & 9.61 & 5.23 & 5.23 \\
\hline
\end{tabular}
\end{table}

\FloatBarrier
\sisubsection{Training data}
\label{si:training-data}
The training set of each system has two parts: grid and random geometries, both generated before any QM calculation, and geometries added by active learning. The validation set and the held-out test set are generated in the same way as the first part.

\paragraph{Grid and random geometries.}
For the \ce{H^{.} + H2} reaction, the geometries are generated from the three H--H distances without any QM calculation: the collinear minimum-energy path of a London--Eyring--Polanyi--Sato surface\cite{Sato1955_NewMethodDrawing} through the saddle point, the antisymmetric and symmetric stretches and the bending at the saddle point, the \ce{H2} potential curve, \ce{H2} at several bond lengths with the third atom beyond the 5.2~\AA\ radial cutoff of the MLP (the asymptote from which the barrier is measured), and random geometries with two H--H distances between 0.45 and 4.5~\AA\ and the angle between them between 40 and 180$^\circ$. Geometries more than 1.5~eV (34.6~kcal/mol) above \ce{H2 + H^{.}} on this surface or without any H--H distance below 1.35~\AA\ are discarded, which keeps the region visited by the reaction at every temperature considered here (the barrier is 0.42~eV and the thermal energy at 2000~K 0.17~eV) and in which single-reference methods, and hence the schemes, are defined.

For ammonia, the geometries are the inversion path, with the N--H distance interpolated linearly between the pyramidal and planar stationary points of the original focal-point analysis,\cite{Csaszar1998_PursuitAbInitioLimit} the symmetric stretch at both stationary points, the umbrella angle at the equilibrium N--H distance, and random displacements of geometries along the path, with every N--H distance between 0.80 and 1.35~\AA. Each system has 4000 training geometries. For the \ce{H^{.} + H2} reaction, 300 further geometries are sampled around the CCSD(T)/cc-pVTZ saddle point from the harmonic quantum Boltzmann distribution at 300 and 1000~K, with the sign of the imaginary frequency flipped, so that the training set describes the curvature of the potential around the saddle point.

\paragraph{Active learning.}
The second part of each training set is selected by active learning with MLatom's standard procedure\cite{MLatom3} (p.~\pageref{si:al}). The MLP of this active learning is trained on one QM method, MP2 with the smaller basis set: MP2/cc-pVTZ for the \ce{H^{.} + H2} reaction and MP2/aug-cc-pVTZ for ammonia. The initial data set has 250 geometries, from Wigner sampling at the minimum, Wigner sampling at the saddle point with the sign of the imaginary frequency flipped, and harmonic quantum Boltzmann sampling at the saddle point. The uncertainty threshold is MLatom's default for the \ce{H^{.} + H2} reaction; for ammonia, it is the median of the default thresholds of five MLPs trained on the initial data set with different validation sets. For the \ce{H^{.} + H2} reaction, the loop is followed by rounds in which the 50 most uncertain geometries of 2.5-ps molecular-dynamics runs are added, until the error on a fixed test set of such geometries stops decreasing; this gives 864 geometries. For ammonia, we use the data set after 11 iterations, 830 geometries, 9 of them added in the last iteration. All these geometries are then calculated with the training QM methods in the same way as the grid and random ones.

\paragraph{Validation set.}

 The validation sets, 274 geometries for the \ce{H^{.} + H2} reaction and 296 for ammonia, were not used in training, but they were used during development: to choose between variants of the training (the extra loss terms for \ce{H^{.} + H2}, the scheme term and the final ensemble for ammonia) and to accept the final ensembles. The errors on them are therefore validation errors, not test errors. They are generated in the same way as the grid and random training geometries: the points of the reaction or inversion path and of the stretches and bends lie halfway between the training points, and the random geometries are drawn with a different random seed. None of them lies within 0.001~\AA\ (root-mean-square displacement per atom in their common generated orientation, without rigid alignment) of a grid/random training geometry, and their median distance to the nearest one is 0.05 (\ce{H^{.} + H2}) and 0.04~\AA\ (ammonia).

\paragraph{Held-out test set.}
For each system, a held-out test set was drawn after the development had finished, and it has not been used for any decision. It is drawn with the random part of the same procedure as the validation set, with a new random seed, and every geometry is at least 0.01~\AA\ (root-mean-square displacement per atom after alignment) from every geometry used before: training, validation, active-learning and other geometries calculated during development. This leaves 204 geometries for the \ce{H^{.} + H2} reaction and 297 for ammonia. They are calculated with the training QM methods and the TBE, with energies only. For the \ce{H^{.} + H2} reaction, four geometries lack an energy of the TBE (calculations not converged) and 11 lie above the energy cap of the training set, so 189 are used for the test.

Geometries more than 34.6~kcal/mol (\ce{H^{.} + H2}) or 42.9~kcal/mol (ammonia) above the reference geometry (p.~\pageref{si:reference-calcs}) with CCSD(T)/cc-pVTZ or CCSD(T)/aug-cc-pVTZ are left out of training and testing: 178 of the 4300 geometries of the \ce{H^{.} + H2} reaction,  and 8 of its 282 validation geometries, leaving 274; with the active-learning geometries (2 of the 864 above the cap for the reaction, 3 of the 830 for ammonia), the MLPs are trained on  4970 (\ce{H^{.} + H2}) and 4827 (ammonia) geometries. The 42.9~kcal/mol ammonia cutoff spans energies well above the inversion barrier. %

\paragraph{QM methods of each geometry.}
 Training uses different numbers of labels for each QM method (Table~\ref{tab:si-training-labels}). We put all geometries in one random order and select a first part of the available labels for each method. Within each basis set, every geometry used to train on a more expensive method is also used to train on the cheaper methods; the subsets need not be nested across basis sets. For the \ce{H^{.} + H2} reaction,  4967 training geometries  are used for training with HF and MP2/cc-pVTZ, 60\% with CCSD and CCSD(T)/cc-pVTZ, 25\% with full CI/cc-pVTZ, 63\% with HF and MP2/cc-pVQZ, and 38\% with CCSD and CCSD(T)/cc-pVQZ. For ammonia, all 4827 training geometries  are used for training with HF and MP2/aug-cc-pVTZ, 60\% with CCSD and CCSD(T)/aug-cc-pVTZ, 50\% with HF and MP2/aug-cc-pVQZ, and 30\% with CCSD and CCSD(T)/aug-cc-pVQZ. The reference geometry (p.~\pageref{si:reference-calcs}) is in the training set of every QM method. %

\FloatBarrier
\sisubsection{Ammonia}
\label{si:extrapolation-discussion}
\label{si:nh3-discussion}

\begin{figure}[htb]
\centering
\includegraphics[width=\linewidth]{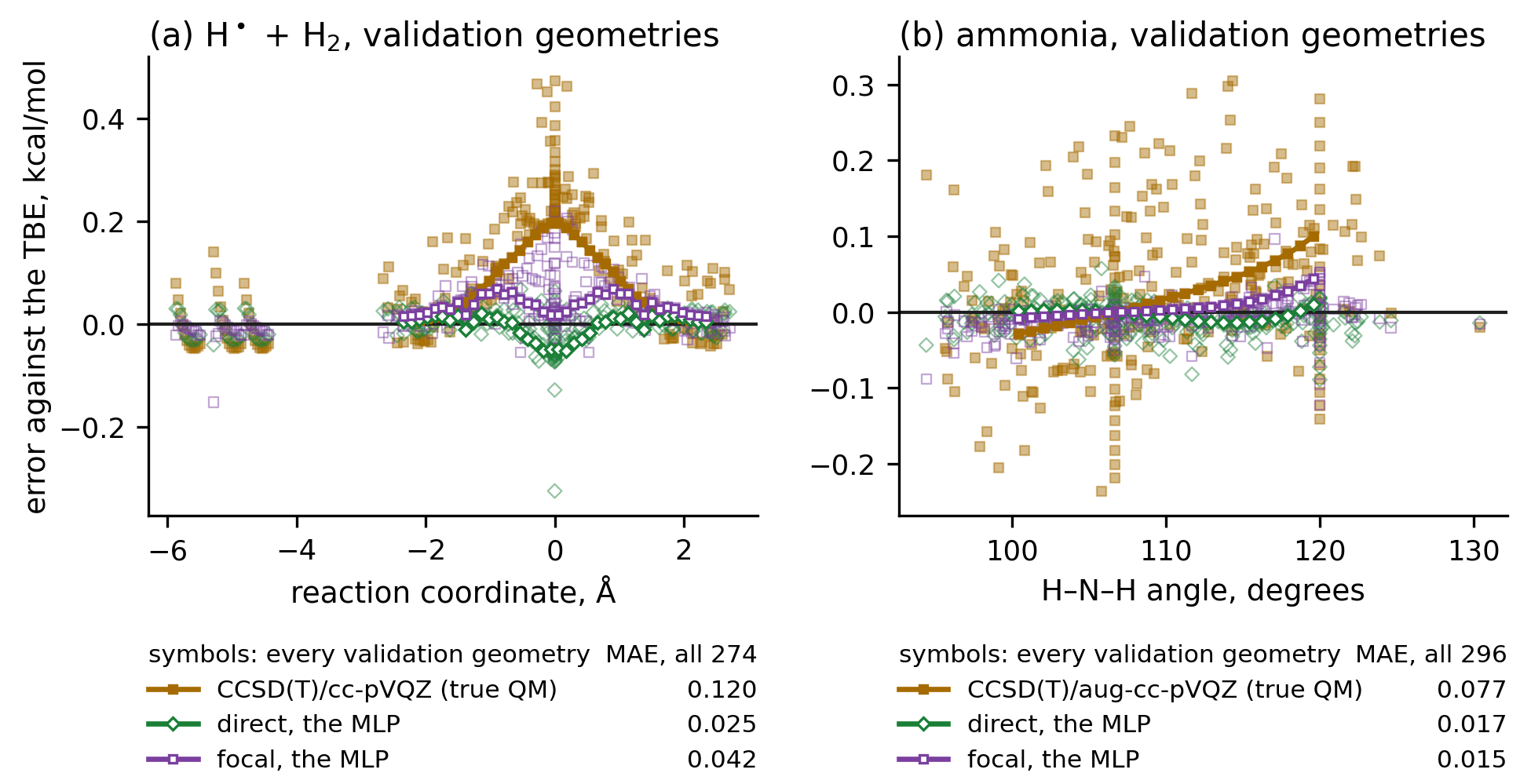}
\caption{\textbf{The MLP over the validation geometries.} Errors against the TBE of the relative energies of the validation geometries (the reaction path, the saddle region, the reactants and thermal distortions around them; lines for the series along the two paths, symbols for every geometry) of the \ce{H^{.} + H2} reaction (a; against the antisymmetric stretch, zero at the saddle point) and of the inversion of ammonia (b; against the H--N--H angle), for the MLP with both schemes and for the most accurate training QM method, CCSD(T) with the larger basis set, in true QM (filled symbols: training QM method; open symbols: unseen schemes); below each panel, the MAE over all validation geometries. The dip of the direct-extrapolation scheme at the saddle point in (a) is that of the scheme itself, which the MLP reproduces ($-0.06$~kcal/mol in true QM).}
\label{fig:heldout}
\end{figure}

\paragraph{Validation set.}
Figure~\ref{fig:heldout} shows the energy errors of the individual validation geometries. On the 179 of the 296 validation ammonia geometries that are farther than 0.01~\AA{} (root-mean-square displacement per atom after alignment) from every training geometry, the mean absolute errors of the focal-point and direct MLPs against the TBE are 0.016 and 0.018~kcal/mol, against 0.010 and 0.012 for the same schemes on true QM and 0.084 for CCSD(T)/aug-cc-pVQZ. On the full set of 296 (Table~\ref{tab:si-nh3-heldout}), focal point has the smaller mean absolute error (0.015 against 0.017~kcal/mol), whereas direct extrapolation is closer to the TBE for the barrier (5.241 against 5.274~kcal/mol; TBE 5.225) and the one-dimensional splitting (1.00 against 0.97 times the TBE). The TBE is the direct scheme with aug-cc-pVQZ and aug-cc-pV5Z, so it shares the aug-cc-pVQZ energies and the form of the extrapolation with the direct scheme; this does not affect the comparison with the training QM methods, but it favours the direct scheme in a comparison of the two schemes with each other. Their ordering also depends on whether a barrier or the surrounding surface is compared (SI, Tables~\ref{tab:si-nh3-splitting} and~\ref{tab:si-nh3-heldout}).

\paragraph{Held-out test set.}
The errors on the held-out test set of ammonia (p.~\pageref{si:training-data}) are 0.017 (focal point) and 0.018~kcal/mol (direct) against the TBE, against 0.010 and 0.013 for the same schemes on true QM and 0.085 for CCSD(T)/aug-cc-pVQZ, the most accurate training QM method (Table~\ref{tab:si-nh3-test}).

\begin{table}[h]
\caption{Held-out test set of the ammonia inversion: 297 geometries never used during development. Relative energies against the TBE, on every surface measured from that surface's own energy at the reference geometry: MAE and maximum absolute error in kcal/mol of the true QM calculations and of the MLP (the ensemble's averaged prediction), and the MAE of the MLP against the same QM method or scheme in true QM.}
\label{tab:si-nh3-test}
\small
\begin{tabular}{lrrrrr}
\hline
 & \multicolumn{2}{c}{true QM} & \multicolumn{2}{c}{the MLP} & MLP vs \\
 & MAE & max & MAE & max & true QM \\
\hline
focal-point scheme & 0.010 & 0.035 & 0.017 & 0.103 & 0.017 \\
direct-extrapolation scheme & 0.013 & 0.071 & 0.018 & 0.070 & 0.015 \\
CCSD(T)/aug-cc-pVQZ & 0.085 & 0.275 & 0.083 & 0.271 & 0.010 \\
MP2/aug-cc-pVTZ & 0.092 & 0.400 & 0.092 & 0.399 & 0.003 \\
CCSD/aug-cc-pVTZ & 0.101 & 0.363 & 0.101 & 0.375 & 0.003 \\
CCSD/aug-cc-pVQZ & 0.133 & 0.730 & 0.133 & 0.706 & 0.008 \\
MP2/aug-cc-pVQZ & 0.171 & 0.723 & 0.171 & 0.720 & 0.005 \\
CCSD(T)/aug-cc-pVTZ & 0.266 & 0.816 & 0.264 & 0.818 & 0.009 \\
HF/aug-cc-pVTZ & 1.227 & 6.068 & 1.229 & 6.070 & 0.003 \\
HF/aug-cc-pVQZ & 1.291 & 6.231 & 1.295 & 6.243 & 0.010 \\
\hline
\end{tabular}
\end{table}

\begin{table}[h]
\caption{Relative energies of the 296 geometries of the validation set (used during development, not a test set) of the ammonia inversion against the TBE, on every surface measured from that surface's own energy at the reference geometry, for the MLP as for the QM methods: MAE and maximum absolute error in kcal/mol of the true QM calculations and of the MLP (the ensemble's averaged prediction), and the MAE of the MLP against the same QM method or scheme in true QM.}
\label{tab:si-nh3-heldout}
\small
\begin{tabular}{lrrrrr}
\hline
 & \multicolumn{2}{c}{true QM} & \multicolumn{2}{c}{the MLP} & MLP vs \\
 & MAE & max & MAE & max & true QM \\
\hline
focal-point scheme & 0.010 & 0.040 & 0.015 & 0.123 & 0.015 \\
direct-extrapolation scheme & 0.011 & 0.069 & 0.017 & 0.089 & 0.013 \\
CCSD(T)/aug-cc-pVQZ & 0.077 & 0.305 & 0.076 & 0.294 & 0.009 \\
MP2/aug-cc-pVTZ & 0.092 & 0.323 & 0.092 & 0.319 & 0.002 \\
CCSD/aug-cc-pVTZ & 0.097 & 0.381 & 0.096 & 0.375 & 0.003 \\
CCSD/aug-cc-pVQZ & 0.118 & 0.639 & 0.118 & 0.622 & 0.007 \\
MP2/aug-cc-pVQZ & 0.154 & 0.630 & 0.154 & 0.652 & 0.004 \\
CCSD(T)/aug-cc-pVTZ & 0.239 & 0.896 & 0.237 & 0.892 & 0.008 \\
HF/aug-cc-pVTZ & 1.091 & 5.346 & 1.092 & 5.351 & 0.003 \\
HF/aug-cc-pVQZ & 1.150 & 5.540 & 1.152 & 5.532 & 0.008 \\
\hline
\end{tabular}
\end{table}

\paragraph{Barrier and surface.}
Figure~\ref{fig:extrapolation} gives the complete method/basis barrier comparison. The barrier alone can conceal errors in a training method. In true QM, CCSD/aug-cc-pVQZ gives 5.226~kcal/mol, close to the 5.225~kcal/mol TBE, because the perturbative-triples correction raises the barrier by 0.105~kcal/mol while the subsequent basis-set correction lowers it by 0.106~kcal/mol. The CCSD/aug-cc-pVQZ profile nevertheless differs from the TBE by 0.505~kcal/mol at 86$^\circ$, and its mean absolute error over the 297 held-out test geometries is 0.133~kcal/mol (Figure~\ref{fig:extrapolation-overview}c; 0.118 on the 296 validation geometries). Agreement for the barrier alone therefore does not establish accuracy over the surrounding surface.

\paragraph{Inversion profile.}
The focal-point MLP follows the relaxed inversion profile more closely than the best training QM method, CCSD(T)/aug-cc-pVQZ, up to about 12~kcal/mol above the minimum. This comparison is limited by geometry: at 88$^\circ$ H--N--H (11.7~kcal/mol on the TBE profile), its error is $-0.085$ against $-0.092$~kcal/mol for CCSD(T)/aug-cc-pVQZ, but at 86$^\circ$ (14.4~kcal/mol) it is $-0.134$ against $-0.104$~kcal/mol (Figure~\ref{fig:extrapolation-overview}e); by 78$^\circ$, outside the panel's range, it reaches $-0.52$~kcal/mol. Only 82 active-learning training geometries lie below 93$^\circ$, down to 86$^\circ$, whereas the same focal-point scheme on true QM energies stays within 0.016~kcal/mol of the TBE over this inner-wall range (p.~\pageref{si:nh3-1d}). Extrapolation in the QM-method space therefore does not establish accuracy at arbitrary geometries.

\paragraph{Inversion splittings.}
On the relaxed inversion path, the one-dimensional splittings of \ce{NH3} on the focal-point and direct MLP surfaces are 0.97 and 1.00 times that on the TBE surface, against 0.98 and 1.01 for the schemes on true QM and 0.89 for the MLP at CCSD(T)/aug-cc-pVQZ (0.90 in true QM; SI, Table~\ref{tab:si-nh3-splitting}). The one-dimensional treatment gives 1.44~cm$^{-1}$ on the TBE surface (p.~\pageref{si:nh3-1d}).

\begin{table}[h]
\caption{Inversion barrier (kcal/mol) and splittings (cm$^{-1}$) of ammonia and its isotopologues on every surface, in true QM and from the MLP (the ensemble), with the same one-dimensional treatment of the relaxed inversion path; in parentheses the ratio to the splitting of the true QM TBE surface, so that every surface is compared under the same treatment. The measured splitting of NH$_3$ is 0.79~cm$^{-1}$.}
\label{tab:si-nh3-splitting}
\small
\resizebox{\linewidth}{!}{%
\begin{tabular}{lrrrrrrrr}
\hline
 & \multicolumn{4}{c}{true QM} & \multicolumn{4}{c}{the MLP} \\
 & barrier & NH$_3$ & ND$_3$ & $^{15}$NH$_3$ & barrier & NH$_3$ & ND$_3$ & $^{15}$NH$_3$ \\
\hline
HF/aug-cc-pVTZ & 4.828 & 3.530 (2.45) & 0.322 (3.29) & 3.377 (2.47) & 4.816 & 3.551 (2.47) & 0.324 (3.31) & 3.397 (2.48) \\
MP2/aug-cc-pVTZ & 5.053 & 1.701 (1.18) & 0.123 (1.26) & 1.620 (1.18) & 5.048 & 1.701 (1.18) & 0.123 (1.26) & 1.621 (1.18) \\
CCSD/aug-cc-pVTZ & 5.419 & 1.287 (0.89) & 0.084 (0.86) & 1.223 (0.89) & 5.410 & 1.289 (0.89) & 0.084 (0.86) & 1.225 (0.89) \\
CCSD(T)/aug-cc-pVTZ & 5.532 & 1.068 (0.74) & 0.066 (0.67) & 1.014 (0.74) & 5.533 & 1.062 (0.74) & 0.065 (0.67) & 1.009 (0.74) \\
HF/aug-cc-pVQZ & 4.785 & 3.687 (2.56) & 0.341 (3.49) & 3.528 (2.58) & 4.778 & 3.688 (2.56) & 0.341 (3.48) & 3.529 (2.58) \\
MP2/aug-cc-pVQZ & 4.884 & 2.019 (1.40) & 0.155 (1.58) & 1.925 (1.41) & 4.892 & 2.008 (1.39) & 0.154 (1.57) & 1.915 (1.40) \\
CCSD/aug-cc-pVQZ & 5.226 & 1.562 (1.08) & 0.109 (1.11) & 1.487 (1.09) & 5.237 & 1.553 (1.08) & 0.108 (1.11) & 1.479 (1.08) \\
CCSD(T)/aug-cc-pVQZ & 5.331 & 1.298 (0.90) & 0.085 (0.87) & 1.234 (0.90) & 5.348 & 1.288 (0.89) & 0.084 (0.86) & 1.225 (0.89) \\
direct-extrapolation scheme & 5.215 & 1.456 (1.01) & 0.099 (1.01) & 1.385 (1.01) & 5.241 & 1.443 (1.00) & 0.098 (1.01) & 1.373 (1.00) \\
focal-point scheme & 5.248 & 1.409 (0.98) & 0.095 (0.97) & 1.340 (0.98) & 5.274 & 1.390 (0.97) & 0.093 (0.95) & 1.322 (0.96) \\
TBE & 5.225 & 1.440 (1.00) & 0.098 (1.00) & 1.370 (1.00) &  &  &  &  \\
\hline
\end{tabular}
}
\end{table}

For comparison with experiment, the splittings are calculated in full dimension with ring-polymer instanton theory\cite{Richardson2011_RingPolymerInstanton} (p.~\pageref{si:instanton-methods}; SI, Table~\ref{tab:si-nh3-instanton}) for \ce{NH3}, \ce{ND3}, and $^{15}$\ce{NH3}, against the measured 0.793,\cite{Urban1984_ForbiddenBandInversionRotation} 0.053,\cite{Fusina1985_InversionRotationSpectrumND3} and 0.758~cm$^{-1}$.\cite{Urban1985_PerturbationAllowedNu4Band15NH3} On the focal-point, direct and CCSD(T)/aug-cc-pVQZ surfaces of the MLP, they are 0.754, 0.764 and 0.669~cm$^{-1}$ for \ce{NH3}, 0.0473, 0.0487 and 0.0411~cm$^{-1}$ for \ce{ND3}, and 0.719, 0.730 and 0.638~cm$^{-1}$ for $^{15}$\ce{NH3}: both schemes raise the splitting of the best training QM method by 13--19\,\%.
The remaining difference from experiment is not a test of the surfaces: they are valence-only Born--Oppenheimer surfaces without core--valence and relativistic corrections, and on one-dimensional ammonia double wells the instanton approximation itself is 7--10\,\% above the exact splitting, while its full-dimensional error for ammonia has not been benchmarked.

\begin{table}[h]
\caption{Full-dimensional inversion splittings (cm$^{-1}$) of ammonia and its isotopologues from ring-polymer instanton theory (half ring polymer of $N$ beads at 25~K, networks in double precision) on the final \ce{NH3} MLP (five-model ensemble trained on the archived and the active-learning geometries, without the errors of the schemes in the training loss) with the direct and focal-point schemes and at its highest training method, CCSD(T)/aug-cc-pVQZ; beside them, experiment. Change with $N$: splitting at $N = 512$ relative to $N = 1024$. On one-dimensional ammonia double wells, the same instanton calculation is 7--10\,\% above the exact splitting.}
\label{tab:si-nh3-instanton}
\small
\begin{tabular}{llrrrr}
\hline
surface & & $N = 1024$ & $N=512$ (\%) & experiment & ratio \\
\hline
direct extrapolation & NH$_3$ & 0.7645 & +1.7 & 0.7934 & 0.96 \\
 & ND$_3$ & 0.04867 & +0.9 & 0.053 & 0.92 \\
 & $^{15}$NH$_3$ & 0.7295 & +1.7 & 0.7577 & 0.96 \\
focal point & NH$_3$ & 0.7539 & +1.7 & 0.7934 & 0.95 \\
 & ND$_3$ & 0.04730 & +0.9 & 0.053 & 0.89 \\
 & $^{15}$NH$_3$ & 0.7192 & +1.7 & 0.7577 & 0.95 \\
CCSD(T)/aug-cc-pVQZ & NH$_3$ & 0.6694 & +1.7 & 0.7934 & 0.84 \\
 & ND$_3$ & 0.04105 & +0.9 & 0.053 & 0.77 \\
 & $^{15}$NH$_3$ & 0.6383 & +1.7 & 0.7577 & 0.84 \\
\hline
\end{tabular}
\end{table}

\paragraph{Cost of the instanton calculation.}
For the reported focal-point \ce{NH3} splitting, optimization of the five-model MLP surface at 64, 256, and 1024 beads, including a continuation of the final stage, recorded 303, 1319, and 4704 steps, respectively. Counting the bead geometries at each step gives about 5.17 million bead energy-and-force evaluations during optimization; these are repeated evaluations, not distinct training geometries. The single run was allocated four CPU cores (Intel Xeon Gold 6240) for 9.83~h (39.3 allocated core-hours). Its quoted splitting uses analytic Hessians at the 1024 final beads; an i-PI finite-difference bead Hessian was also evaluated as a check. A matched direct-QM runtime is unavailable: evaluating the focal-point combination at every bead would require its energy and gradient components, including CCSD(T)/aug-cc-pVTZ, whose analytic gradient in the PySCF version used here failed the finite-difference check (p.~\pageref{si:reference-calcs}).

\FloatBarrier
\sisubsection{Hydrogen exchange: extrapolation to full CI/CBS}
\label{si:h3-discussion}

The same construction reaches the exact solution when full CI is affordable. For the \ce{H^{.} + H2 -> H2 + H^{.}} reaction, whose exact Born--Oppenheimer barrier is known from quantum Monte Carlo calculations, 9.6089~$\pm$~0.0001~kcal/mol,\cite{Anderson2016_ExactBarrierHH2,Diedrich1992_AccurateQuantumChemistry,Riley2003_HigherAccuracyQuantum,Mielke2005_BornOppenheimerDiagonalCorrection} we train one MLP on HF, MP2, CCSD, and CCSD(T) with cc-pVTZ and cc-pVQZ and on full CI/cc-pVTZ, and add $\Delta E_{\mathrm{FCI}}$ (Eq.~\ref{eq:increments}) to both schemes, so that both target full CI/CBS (p.~\pageref{si:schemes}). Even full CI, exact for a given basis set, overestimates the barrier by 0.1~kcal/mol with cc-pVQZ and by 0.3 with cc-pVTZ (SI, Figure~\ref{fig:extrapolation}a). The TBE is the direct scheme with cc-pVQZ and cc-pV5Z plus $\Delta E_{\mathrm{FCI}}$; its barrier, 9.59~kcal/mol, is within 0.02~kcal/mol of the exact one.

The direct and focal-point MLP barriers are 9.54 and 9.61~kcal/mol, against 9.53 and 9.60~kcal/mol for the same schemes on true QM and 9.79~kcal/mol for the best training QM method, CCSD(T)/cc-pVQZ (SI, Table~\ref{tab:si-h3-saddle}). On the 274 validation geometries, the mean absolute errors against the TBE are 0.025 and 0.042~kcal/mol, against 0.023 and 0.037 for the schemes on true QM and 0.120 for CCSD(T)/cc-pVQZ (Table~\ref{tab:si-h3-heldout}); on the 237 of them farther than 0.01~\AA{} from every training geometry, 0.023 and 0.044, against 0.021, 0.039 and 0.114. The errors on the held-out test set of this reaction (p.~\pageref{si:training-data}) are 0.022 (direct) and 0.059~kcal/mol (focal point) against the TBE, against 0.019 and 0.047 for the same schemes on true QM and 0.118 for CCSD(T)/cc-pVQZ (Table~\ref{tab:si-h3-test}): the focal-point MLP is two times more accurate than the most accurate training QM method, short of the margin of 0.05~kcal/mol at the upper 95\,\% confidence bound that we set before the calculation, as on the 132 random geometries of the validation set that are at least 0.01~\AA{} (root-mean-square displacement per atom after alignment) from every training geometry (0.052 against 0.109~kcal/mol).

\begin{table}[h]
\caption{Held-out test set of the \ce{H^{.} + H2} reaction: 189 geometries never used during development. Relative energies against the TBE, on every surface measured from that surface's own energy at the reference geometry: MAE and maximum absolute error in kcal/mol of the true QM calculations and of the MLP (the ensemble's averaged prediction), and the MAE of the MLP against the same QM method or scheme in true QM.}
\label{tab:si-h3-test}
\small
\begin{tabular}{lrrrrr}
\hline
 & \multicolumn{2}{c}{true QM} & \multicolumn{2}{c}{the MLP} & MLP vs \\
 & MAE & max & MAE & max & true QM \\
\hline
direct-extrapolation scheme & 0.019 & 0.070 & 0.022 & 0.410 & 0.014 \\
focal-point scheme & 0.047 & 0.198 & 0.059 & 0.292 & 0.018 \\
CCSD(T)/cc-pVQZ & 0.118 & 0.483 & 0.116 & 0.458 & 0.013 \\
CCSD/cc-pVQZ & 0.184 & 0.790 & 0.185 & 0.857 & 0.014 \\
full CI/cc-pVTZ & 0.261 & 1.062 & 0.261 & 1.060 & 0.013 \\
CCSD(T)/cc-pVTZ & 0.277 & 1.118 & 0.274 & 1.096 & 0.012 \\
CCSD/cc-pVTZ & 0.333 & 1.370 & 0.337 & 1.413 & 0.017 \\
MP2/cc-pVQZ & 1.664 & 6.846 & 1.661 & 6.852 & 0.013 \\
MP2/cc-pVTZ & 1.713 & 6.920 & 1.708 & 6.914 & 0.012 \\
HF/cc-pVTZ & 3.004 & 9.432 & 3.001 & 9.456 & 0.011 \\
HF/cc-pVQZ & 3.032 & 9.231 & 3.022 & 9.233 & 0.014 \\
\hline
\end{tabular}
\end{table}

\begin{table}[h]
\caption{Relative energies of the 274 geometries of the validation set (used during development, not a test set) of the \ce{H^{.} + H2} reaction against the TBE, on every surface measured from that surface's own energy at the reference geometry, for the MLP as for the QM methods: MAE and maximum absolute error in kcal/mol of the true QM calculations and of the MLP (the ensemble's averaged prediction), and the MAE of the MLP against the same QM method or scheme in true QM.}
\label{tab:si-h3-heldout}
\small
\begin{tabular}{lrrrrr}
\hline
 & \multicolumn{2}{c}{true QM} & \multicolumn{2}{c}{the MLP} & MLP vs \\
 & MAE & max & MAE & max & true QM \\
\hline
direct-extrapolation scheme & 0.023 & 0.077 & 0.025 & 0.325 & 0.011 \\
focal-point scheme & 0.037 & 0.288 & 0.042 & 0.219 & 0.014 \\
CCSD(T)/cc-pVQZ & 0.120 & 0.472 & 0.117 & 0.462 & 0.011 \\
CCSD/cc-pVQZ & 0.212 & 0.823 & 0.212 & 0.861 & 0.013 \\
full CI/cc-pVTZ & 0.261 & 1.028 & 0.261 & 1.067 & 0.010 \\
CCSD(T)/cc-pVTZ & 0.282 & 1.103 & 0.279 & 1.103 & 0.010 \\
CCSD/cc-pVTZ & 0.363 & 1.413 & 0.368 & 1.425 & 0.014 \\
MP2/cc-pVQZ & 1.660 & 8.057 & 1.657 & 8.125 & 0.012 \\
MP2/cc-pVTZ & 1.709 & 8.100 & 1.706 & 8.157 & 0.010 \\
HF/cc-pVTZ & 3.234 & 9.766 & 3.231 & 9.792 & 0.010 \\
HF/cc-pVQZ & 3.267 & 9.564 & 3.257 & 9.574 & 0.014 \\
\hline
\end{tabular}
\end{table}
\begin{table}[h]
\caption{The collinear saddle point of every surface of the \ce{H^{.} + H2} reaction, in true QM and of the MLP (the ensemble, from its energies at the same geometries): barrier in kcal/mol, the imaginary and the bending frequency of \ce{H-H-H} in cm$^{-1}$, and the rate constant of \ce{D^{.} + H2 -> HD + H^{.}} from the same transition-state theory with Eckart tunneling as the ratio to that of the TBE surface at 300 and 1000~K; the same treatment is used for every surface, with its residual effect on the ratios assessed in Methods.}
\label{tab:si-h3-saddle}
\small
\resizebox{\linewidth}{!}{%
\begin{tabular}{lrrrrrrrr}
\hline
 & \multicolumn{4}{c}{true QM} & \multicolumn{4}{c}{the MLP} \\
 & barrier & $\nu^{\ddagger}$ & bend & $k/k_{\mathrm{TBE}}$ (300, 500, 1000~K) & barrier & $\nu^{\ddagger}$ & bend & $k/k_{\mathrm{TBE}}$ (300, 1000~K) \\
\hline
HF/cc-pVTZ & 17.58 & 2298i & 914 & 0.00, 0.00, 0.03 & 17.56 & 2305i & 940 & 0.00, 0.02 \\
MP2/cc-pVTZ & 13.23 & 1968i & 930 & 0.02, 0.04, 0.16 & 13.22 & 1988i & 929 & 0.02, 0.16 \\
CCSD/cc-pVTZ & 10.30 & 1539i & 917 & 0.31, 0.46, 0.64 & 10.29 & 1540i & 955 & 0.26, 0.58 \\
CCSD(T)/cc-pVTZ & 10.01 & 1510i & 917 & 0.45, 0.59, 0.74 & 10.01 & 1521i & 919 & 0.45, 0.73 \\
HF/cc-pVQZ & 17.63 & 2302i & 903 & 0.00, 0.00, 0.03 & 17.60 & 2311i & 925 & 0.00, 0.02 \\
MP2/cc-pVQZ & 13.04 & 1964i & 905 & 0.02, 0.05, 0.19 & 13.04 & 1978i & 892 & 0.03, 0.20 \\
CCSD/cc-pVQZ & 10.10 & 1536i & 892 & 0.47, 0.60, 0.75 & 10.11 & 1553i & 912 & 0.44, 0.71 \\
CCSD(T)/cc-pVQZ & 9.79 & 1506i & 892 & 0.70, 0.79, 0.88 & 9.78 & 1506i & 894 & 0.68, 0.87 \\
full CI/cc-pVTZ & 9.94 & 1504i & 917 & 0.50, 0.63, 0.76 & 9.95 & 1520i & 919 & 0.50, 0.76 \\
direct-extrapolation scheme & 9.53 & 1493i & 879 & 1.09, 1.06, 1.03 & 9.54 & 1489i & 883 & 1.01, 1.00 \\
focal-point scheme & 9.60 & 1490i & 883 & 0.94, 0.97, 0.98 & 9.61 & 1503i & 867 & 1.01, 1.01 \\
TBE & 9.59 & 1497i & 880 & 1.00, 1.00, 1.00 &  &  &  &  \\
\hline
\end{tabular}
}
\end{table}

\paragraph{Rate constants.}
For \ce{D^{.} + H2 -> HD + H^{.}}, which runs on the same surface, the rate constants from conventional transition-state theory with an Eckart tunneling correction on the direct and focal-point MLP surfaces are 1.01 and 1.00 (direct, 300 and 1000~K) and 1.01 and 1.01 (focal point) times those on the TBE surface, against 1.09 and 1.03, and 0.94 and 0.98 for the same schemes on true QM, and 0.68 and 0.87 for the MLP at CCSD(T)/cc-pVQZ (0.70 and 0.88 in true QM; SI, Table~\ref{tab:si-h3-saddle}). At 200~K, the sensitivity to the tunneling treatment prevents a resolved comparison.

\FloatBarrier
\sisubsection{The errors of the schemes in the training loss}
\label{si:scheme-term-discussion}

The schemes amplify the changes of the energies between the two basis sets (Eq.~\ref{eq:increments}), so a small error of the MLP in such a change is magnified in the extrapolated energy. Adding the errors of both schemes to the training loss makes the MLP precise in exactly these combinations, while keeping both schemes available after training (below). It is an option, not a requirement of the approach.

\paragraph{Hydrogen exchange.}
For the \ce{H^{.} + H2} reaction (p.~\pageref{si:h3-discussion}), the training loss also contains the errors of both schemes, written as the linear combinations of Eq.~\ref{eq:increments}: at every step, the mean squared difference between each scheme applied to the energies predicted by the MLP and to the reference energies, over 16 random training geometries that carry every QM method the schemes combine, with a coefficient of 1. Both schemes are included, so that either can still be chosen after training. Without this term the direct-extrapolation scheme on the MLP was 0.10~kcal/mol from the same scheme on true QM, because the MLP learned the change of the CCSD(T) energy between the two basis sets less accurately than that of MP2, which comes from all geometries and with forces; with it, 0.02~kcal/mol.

For the \ce{H^{.} + H2} reaction, the errors of the energies of CCSD(T) and full CI, which are learned from energies only and from the fewest geometries, are weighted by 3, and those of the reference geometry by 10:  with both the scheme and Hessian terms but without these weights, the combinations were accurate but the full CI error rose to 0.04~kcal/mol, while for ammonia every QM method stays within 0.01~kcal/mol without them. For the \ce{H^{.} + H2} reaction, whose rate constants depend on the curvature at the saddle point, the loss also contains the Hessians of every training QM method at its saddle point and at the reactants, obtained from finite differences of its energies (rate constants, p.~\pageref{si:h3-discussion}), as a mean squared error with a coefficient of 0.01 once per epoch. With these Hessians the MLP's error in the bending frequency at the saddle point is 1.5\% on average over the eleven surfaces, and in the imaginary one 0.6\% (Table~\ref{tab:si-h3-saddle}).

\paragraph{Ammonia.}
For ammonia, the MLP of the main text is trained without it: on the 296 validation geometries, focal point and direct extrapolation on that MLP are 0.015 and 0.017~kcal/mol from the TBE and 0.015 and 0.013 from the same schemes on true QM; an MLP trained on the same data with the scheme term gives 0.011 and 0.011 against the TBE and 0.004 and 0.005 against true QM, with  barriers differing by less than 0.015~kcal/mol and the same one-dimensional splitting (within 1\,\%). On the 297 geometries of the held-out test set, the two schemes on the MLP with the scheme term are 0.010 and 0.013~kcal/mol from the TBE, against 0.017 and 0.018 without it. Both are averages over five-model ensembles trained on the same folds; single models without the term scatter more (standard deviation of the focal-point error over the five models 0.005 against 0.001~kcal/mol), so the comparison holds for ensembles, not necessarily for a single model.

\FloatBarrier
\clearpage
\sisection{Interpolation along exact exchange}
\label{si:coordinate}
\label{si:methods-interpolation}

\sisubsection{Training data and active learning}
\label{si:coordinate-al}

For malonaldehyde, the active learning (p.~\pageref{si:al}) uses the MLP with the method descriptor $a$, and every geometry it selects is calculated with all three training functionals. The uncertainty threshold is one value for all three trained values of $a$; it is determined once, on the initial data set, and kept for the rest of the loop.

The initial data set is built with MLatom's procedure:\cite{AL} geometries are added 50 at a time until the learning curve of the five-fold cross-validated energy error predicts less than 10\,\% improvement from 50 more geometries. The geometries come in equal numbers from nine samplings at 300~K: for each training functional, Wigner sampling at its minimum, Wigner sampling at its transition state with the sign of the imaginary frequency flipped, and harmonic quantum Boltzmann sampling at its transition state. Two loops were run. The first started from MLatom's own initial data set of 350 geometries, with its uncertainty threshold fitted on that set (2.57~kcal/mol), and was stopped after five iterations at 965 geometries. The second started from the first 250 of these geometries, with a fixed threshold, the median of five fits on them (4.09~kcal/mol), and was stopped after seven iterations at 1111 geometries. The final data set is their union, 1826 geometries after removing the 250 shared ones. Neither loop had met the convergence criterion below when stopped. The union was chosen over the data of either loop because its MLP stays intact and free of energy holes in more of the production trajectories (p.~\pageref{si:coordinate-checks}); these trajectories selected the model rather than testing it.

In every iteration, the MLP runs the following simulations at each of the three trained values of $a$: four geometry optimizations of the minimum and four of the transition state with geomeTRIC,\cite{Wang2016_GeometryOptimizationMadeSimple} started from the samples of the initial data set; the IRC from the transition state of the MLP; and molecular dynamics in the NVE ensemble for 2.5~ps with a time step of 0.5~fs. Sixteen trajectories start from Wigner sampling at the minimum and sixteen at the transition state of the MLP: eight from harmonic quantum Boltzmann sampling with a thermal momentum of random sign along the reaction coordinate, and eight from Wigner sampling with the sign of the imaginary frequency flipped.

A trajectory stops at its first uncertain geometry; an optimization or an IRC runs to the end, and its first uncertain geometry is taken. If an iteration finds fewer uncertain geometries than it has simulations, the simulations are repeated from new initial conditions, up to 25 times. The loop has converged when fewer than 5\% of the simulations of an iteration find a new uncertain geometry.\cite{AL} The geometries of the data set are calculated only with the three training functionals; the two functionals not used in training, BLYP35 and BHandHLYP, are calculated only for the test geometries.

\FloatBarrier
\sisubsection{The MLP and the choice of $a_\mathrm{opt}$}
\label{si:coordinate-mlp}

\paragraph{The model.}
The MLP extends the OMNI-P models from a one-hot QM-method descriptor to a continuous fraction of exact exchange, $a$, alongside the nuclear coordinates.\cite{AIO-OMNI-P1,OMNIP2x} Its energy is a sum of atomic energies; the network for each element receives that atom's chemical environment and $a$ (p.~\pageref{si:mlp}). The three training functionals are BLYP ($a=0$), B1LYP ($a=0.25$), and HFLYP ($a=1$). Energies and forces were calculated with cc-pVDZ for 1826 geometries generated by active learning (p.~\pageref{si:coordinate-al}).  The malonaldehyde proton-transfer barrier is sensitive to exact exchange: the barrier-oriented MPW1K and BB1K functionals contain 42.8\% and 42\% exact exchange, respectively, compared with 20\% for B3LYP.\cite{Lynch2000_AdiabaticConnectionKinetics,Zhao2004_DevelopmentAssessmentNew} BLYP35 ($a=0.35$) and BHandHLYP ($a=0.5$) are withheld named members of the same functional family; the family can also be extended to multiple parameters, as in B3LYP-type hybrids.\cite{Renz2009_ReliableQuantumChemical,Stephens1994_AbInitioCalculationVibrational}

\paragraph{The reported MLP and $a_\mathrm{opt}$.}
All reported results are obtained with an ensemble of five MLPs trained on the 1826 geometries of the final data set with different random seeds and training--validation splits, whose predictions are averaged, as in our latest work on vibrational frequencies.\cite{ANI-1ccx-gelu-anharmonic} The networks are evaluated in double precision for optimizations, Hessians and instantons. The optimal fraction $a_\mathrm{opt}$ is the one at which the barrier of the MLP at the two B3LYP stationary geometries equals the CCSD(T) barrier there, found by bisection on the MLP after a scan over 101 values of $a$ between 0 and 1.

\paragraph{Selection of $a_\mathrm{opt}$.}
The selected $a_\mathrm{opt}=0.455$ matches the CCSD(T)/cc-pVDZ barrier at two fixed B3LYP/cc-pVDZ stationary geometries using two CCSD(T) single-point energies (Figure~\ref{fig:coordinate}d). This matched barrier selects the coordinate; it is not an independent validation of the resulting MLP. At each method's own relaxed minimum and transition state, the training functionals miss the CCSD(T) reference by 1.23 (B1LYP) to 3.33 (HFLYP)~kcal/mol, while the optimal MLP misses it by 0.05~kcal/mol (Table~\ref{tab:si-coordinate-relaxed}).

\begin{table}[h]
\caption{Relaxed barriers in kcal/mol: each method's barrier between its own minimum and transition state, the CCSD(T) barrier in the same basis at the same two geometries, and the difference.}
\label{tab:si-coordinate-relaxed}
\begin{tabular}{lccc}
\hline
 & \multicolumn{3}{c}{cc-pVDZ} \\
method & own & CCSD(T) there & difference \\
\hline
BLYP & 1.20 & 3.57 & $-2.36$ \\
B1LYP & 2.65 & 3.88 & $-1.23$ \\
BLYP35 & 3.28 & 3.91 & $-0.62$ \\
BHandHLYP & 4.23 & 3.87 & $+0.36$ \\
HFLYP & 6.96 & 3.64 & $+3.33$ \\
MLP at $a_\mathrm{opt}$ & 3.94 & 3.90 & $+0.05$ \\
\hline
\end{tabular}
\end{table}

\FloatBarrier
\sisubsection{Checks along the reaction path and across the method space}
\label{si:coordinate-checks}
We check the MLP in several regions of the potential energy surface, because no single test is enough to judge its quality and each test probes a different part of the surface: the IRC between the minimum and the transition state, the barrier of every method between its own relaxed minimum and transition state, molecular dynamics at 300~K, and frequencies, zero-point energies and tunneling splittings across $a$. Different tests can rank the methods differently: BHandHLYP, for example, has the smallest error along the IRC (Table~\ref{tab:si-coordinate-path}) but an error of 0.36~kcal/mol in its relaxed barrier, against 0.05~kcal/mol for the MLP at $a_\mathrm{opt}$ (Table~\ref{tab:si-coordinate-relaxed}).

\paragraph{Validation path.}
The MLP at $a_\mathrm{opt}$ is validated along the B3LYP/cc-pVDZ IRC (p.~\pageref{si:reference-calcs}), leaving out the transition state and the two ends of the path, which are or are near the geometries $a_\mathrm{opt}$ is chosen on. The two halves of the path are mirror images, so its 24 interior points are 12 distinct geometries. The validation geometries are therefore the same for every method compared, and none of them is chosen by the MLP (Table~\ref{tab:si-coordinate-path}). On the held-out test set, 132 geometries near this path never used during development, the MLP at $a_\mathrm{opt}$ has a mean absolute error of 0.43~kcal/mol against CCSD(T), and B1LYP, the most accurate training functional, 0.51; the 95\,\% confidence interval of the difference, $-0.28$ to $+0.07$~kcal/mol (bootstrap over IRC points), includes zero (Table~\ref{tab:si-coordinate-test}; the bootstrap resamples the 12 path points with their 11 displaced geometries each).

\begin{table}[h]
\caption{Errors of every method against CCSD(T)/cc-pVDZ along the B3LYP/cc-pVDZ IRC, the validation set (used during development): mean absolute error (MAE) in kcal/mol of the energies of the interior points of the IRC (the transition state and the two ends left out), relative to the same method at the B3LYP minimum. The geometries are the same for every method and none is chosen by the MLP. The two halves of the IRC are mirror images, so its 24 interior points are 12 symmetry-unique geometries. The two null models need the DFT energies of the trained functionals at every geometry.}
\label{tab:si-coordinate-path}
\small
\begin{tabular}{llc}
\hline
method & role & MAE \\
\hline
BHandHLYP & not trained & 0.130 \\
MLP at $a_\mathrm{opt}$ &  & 0.138 \\
real functional at $a_\mathrm{opt}$ &  & 0.147 \\
parabola in $a$ at $a_\mathrm{opt}$ & null model & 0.168 \\
line B1LYP--HFLYP at $a_\mathrm{opt}$ & null model & 0.200 \\
BLYP35 & not trained & 0.303 \\
B1LYP & trained & 0.455 \\
HFLYP & trained & 0.478 \\
B3LYP & comparator & 0.631 \\
BLYP & trained & 0.838 \\
\hline
\end{tabular}
\end{table}

\begin{table}[h]
\caption{Held-out test set of malonaldehyde: 132 geometries never used during development, displaced from the 12 symmetry-unique interior points of the B3LYP/cc-pVDZ IRC by classical normal-mode motion at 50~K. Mean absolute error (MAE) in kcal/mol against CCSD(T)/cc-pVDZ of the energies relative to the same method at the B3LYP minimum.}
\label{tab:si-coordinate-test}
\small
\begin{tabular}{llc}
\hline
method & role & MAE \\
\hline
BLYP35 & not trained & 0.409 \\
line B1LYP--HFLYP at $a_\mathrm{opt}$ & null model & 0.424 \\
MLP at $a_\mathrm{opt}$ &  & 0.425 \\
real functional at $a_\mathrm{opt}$ &  & 0.428 \\
parabola in $a$ at $a_\mathrm{opt}$ & null model & 0.433 \\
BHandHLYP & not trained & 0.468 \\
B1LYP & trained & 0.514 \\
B3LYP & comparator & 0.696 \\
BLYP & trained & 0.969 \\
HFLYP & trained & 1.234 \\
\hline
\end{tabular}
\end{table}

Each method is also judged at its own stationary points: its barrier between its own relaxed minimum and transition state against CCSD(T) at the same two geometries. Two null models, which need DFT at every geometry, are compared as well: the straight line in $a$ between the two trained functionals on either side of $a_\mathrm{opt}$, and the parabola through all three trained functionals.

\paragraph{Checks along the IRC.}

DFT with the functional at $a_\mathrm{opt}$, which is not a training functional, shows how accurately the MLP interpolates in $a$: the MLP differs from it by 0.01~kcal/mol on the barrier and 0.010~kcal/mol on average along the B3LYP IRC (Tables~\ref{tab:si-coordinate-production-dz} and \ref{tab:si-coordinate-fit}). The true functional meets the CCSD(T) barrier at $a=0.455$ (Figure~\ref{fig:coordinate}d). Table~\ref{tab:si-coordinate-barriers} gives the test against CCSD(T) at the B3LYP stationary geometries. At the 12 symmetry-unique IRC geometries, the MLP at $a_\mathrm{opt}$ has a mean absolute error of 0.138~kcal/mol against CCSD(T), versus 0.455--0.838 for its three training functionals and 0.631 for B3LYP at the same geometries (Figure~\ref{fig:coordinate}e; Table~\ref{tab:si-coordinate-path}). A straight line in $a$ between the adjacent training functionals gives 0.200~kcal/mol and a parabola through all three gives 0.168~kcal/mol; both require DFT at every geometry.

The MLP's errors against the training functionals along the IRC are 0.007--0.015~kcal/mol, and against withheld BHandHLYP and BLYP35 are 0.009 and 0.013~kcal/mol (Table~\ref{tab:si-coordinate-fit}). The actual functional at $a_\mathrm{opt}$ has an error against CCSD(T) along the IRC of 0.147~kcal/mol; the nearby withheld BHandHLYP gives 0.130~kcal/mol (Table~\ref{tab:si-coordinate-path}). These checks go beyond the two energies used to select $a$; they establish accuracy along the IRC, rather than across all configurations or compounds.

\begin{table}[h]
\caption{The MLP against DFT with the same functional along the B3LYP/cc-pVDZ IRC: mean absolute difference (MAD) in kcal/mol between the MLP at a value of $a$ and DFT with the functional of the same $a$, over the 12 symmetry-unique interior geometries, for the three training functionals, for the two functionals not used in training (BLYP35 and BHandHLYP; the MLP interpolates in $a$ between the training functionals), and for the functional with $a = a_\mathrm{opt}$. The errors of the MLP against CCSD(T) are in Table~\ref{tab:si-coordinate-path}.}
\label{tab:si-coordinate-fit}
\begin{tabular}{llc}
\hline
functional & & MAD \\
\hline
BLYP & fit & 0.009 \\
B1LYP & fit & 0.015 \\
BLYP35 & interpolation & 0.013 \\
BHandHLYP & interpolation & 0.009 \\
HFLYP & fit & 0.007 \\
real functional at $a_\mathrm{opt}$ & at $a_\mathrm{opt}$ & 0.010 \\
\hline
\end{tabular}
\end{table}
\begin{table}[h]
\caption{The barrier (kcal/mol) at the two B3LYP/cc-pVDZ stationary geometries: the five named functionals of Eq.~\ref{eq:line}, B3LYP and CCSD(T), each functional's difference from CCSD(T), and the fraction of exact exchange at which the real functional meets CCSD(T) (linear interpolation on a grid of 0.05 in $a$).}
\label{tab:si-coordinate-barriers}
\begin{tabular}{lccc}
\hline
 & & \multicolumn{2}{c}{cc-pVDZ} \\
method & $a$ & barrier & $-$CCSD(T) \\
\hline
BLYP & 0.00 & 1.16 & $-2.64$ \\
B1LYP & 0.25 & 2.63 & $-1.16$ \\
BLYP35 & 0.35 & 3.21 & $-0.59$ \\
BHandHLYP & 0.50 & 4.04 & +0.25 \\
HFLYP & 1.00 & 6.33 & +2.53 \\
B3LYP & -- & 2.15 & $-1.65$ \\
CCSD(T) & -- & 3.80 & -- \\
real functional = CCSD(T) & & \multicolumn{2}{c}{$a = 0.455$} \\
\hline
\end{tabular}
\end{table}

\begin{figure}[h]
\centering
\includegraphics[width=\linewidth]{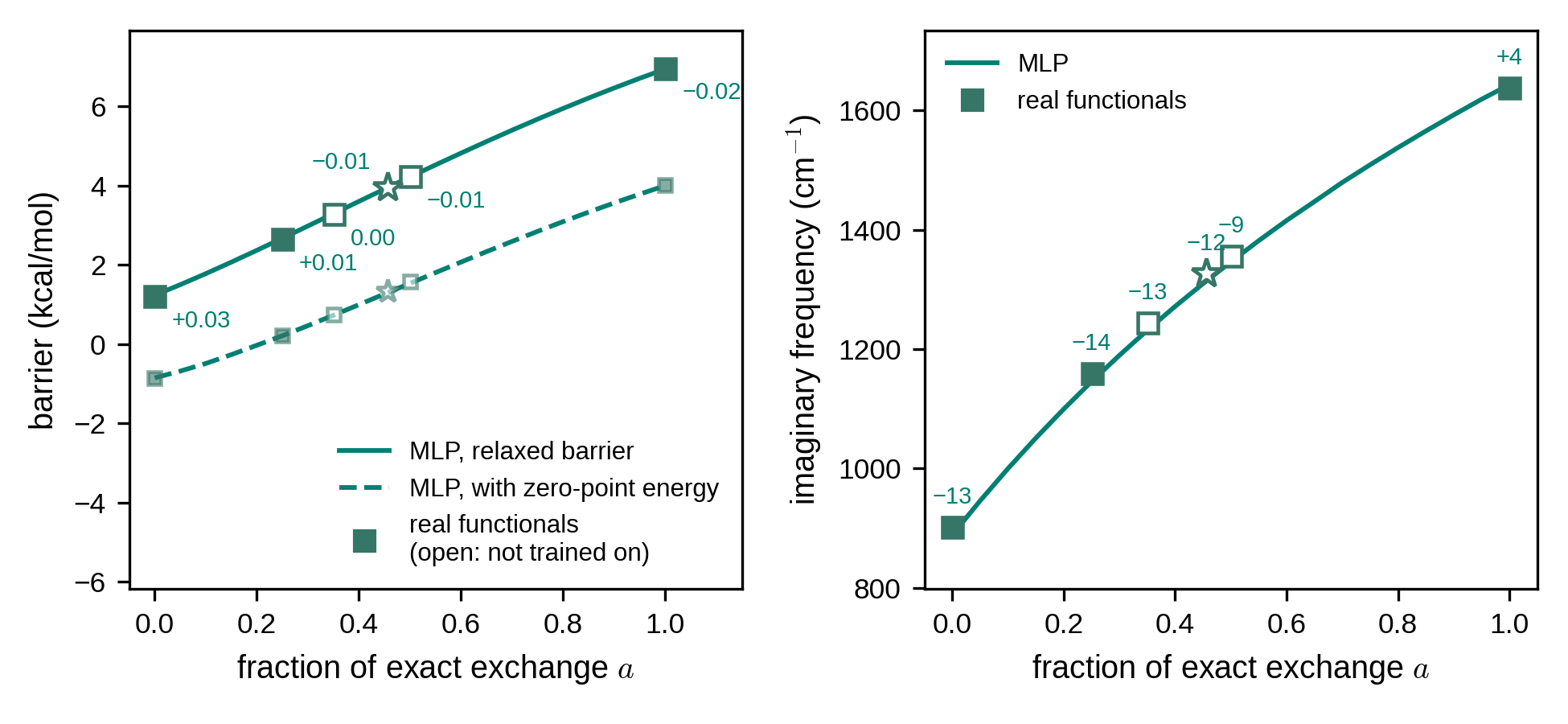}
\caption{The MLP across the space (cc-pVDZ): the relaxed barrier between its own minimum and transition state, with and without the harmonic zero-point energy, and the imaginary frequency at the transition state, on a grid of 21 values of $a$ (lines), against the real functionals at their own stationary points (symbols; the star is the real functional at $a_\mathrm{opt}$); the numbers are the MLP minus the real functional.}
\label{fig:si-coordinate-production-dz}
\end{figure}

\paragraph{Properties across the space.}
The same MLP gives relaxed reaction profiles, harmonic frequencies, zero-point energies and tunneling splittings across $a$. Relative to true DFT at the five named functionals, its relaxed barriers differ by at most 0.03~kcal/mol and its harmonic frequencies at the minimum by 1.8--3.2~cm$^{-1}$ on average (Figure~\ref{fig:si-coordinate-production-dz}; Table~\ref{tab:si-coordinate-production-dz}). At $a_\mathrm{opt}$, including zero-point energy lowers the 3.94~kcal/mol barrier to 1.29~kcal/mol. In 2.5-ps production dynamics at 300~K, 50, 49 and 45 of 50 trajectories at $a = 0$, 0.25 and 1 and 47 of 50 at $a_\mathrm{opt}$ remain intact and free of energy holes at the recomputed snapshots; ensemble disagreement flags every trajectory that breaks or enters a hole before it does. Production dynamics: 50 NVE trajectories of 2.5~ps per value of $a$ (time step 0.5~fs) from 300~K Wigner samples of the MLP's own minimum, on the ensemble average. A trajectory breaks when a hydrogen atom is more than 2~\AA{} from every heavy atom, the heavy atoms are no longer connected (1.3 times the covalent radii), or two atoms come closer than 0.6~\AA. Snapshots every 100~fs, and those of largest ensemble disagreement, most extreme bond angle and largest distance from the minimum, are recomputed with the true functional at that $a$; a trajectory enters an energy hole if the MLP differs from it there by more than 10~kcal/mol, or lies below the B3LYP/cc-pVDZ minimum, the zero of both, while the functional is more than 20~kcal/mol above it. Before the first hole of each such trajectory, every saved snapshot (10~fs apart) is recomputed, to time the hole against the flag. Holes between recomputed snapshots are not seen, so the hole-free counts are upper bounds. The disagreement is the largest deviation of a member from the average; its threshold is the first loop's, 2.57~kcal/mol.

\begin{table}[h]
\caption{Relaxed barrier (kcal/mol), barrier with the harmonic zero-point energy, and imaginary frequency (cm$^{-1}$) at each method's own stationary points, from the real functional and from the MLP at the same $a$ (cc-pVDZ), and the mean absolute difference of the harmonic frequencies of the minimum between the two (cm$^{-1}$).}
\label{tab:si-coordinate-production-dz}
\begin{tabular}{lccccccccc}
\hline
 & & \multicolumn{3}{c}{real functional} & \multicolumn{3}{c}{MLP} & \\
method & $a$ & barrier & +ZPE & $|\omega_\mathrm{TS}|$ & barrier & +ZPE & $|\omega_\mathrm{TS}|$ & frequencies \\
\hline
BLYP & 0.000 & 1.20 & $-0.87$ & 902 & 1.23 & $-0.85$ & 889 & 3.2 \\
B1LYP & 0.250 & 2.65 & 0.21 & 1159 & 2.67 & 0.21 & 1146 & 2.0 \\
BLYP35 & 0.350 & 3.28 & 0.74 & 1245 & 3.29 & 0.73 & 1232 & 2.1 \\
BHandHLYP & 0.500 & 4.23 & 1.57 & 1356 & 4.22 & 1.53 & 1346 & 1.8 \\
HFLYP & 1.000 & 6.96 & 4.00 & 1638 & 6.94 & 4.00 & 1643 & 2.6 \\
$a_\mathrm{opt}$ & 0.455 & 3.95 & 1.32 & 1326 & 3.94 & 1.29 & 1314 & 2.0 \\
B3LYP & -- & 2.15 & $-0.21$ & 1088 & -- & -- & -- & -- \\
\hline
\end{tabular}
\end{table}

\FloatBarrier
\sisubsection{Tunneling splittings of malonaldehyde}
\label{si:coordinate-discussion}
\paragraph{Calculation.}
The splittings are calculated with the ring-polymer instanton protocol of p.~\pageref{si:instanton-methods} on the MLP at a given $a$. The minima are optimized with the tightest criteria of geomeTRIC. For the deuterated molecule, the transferring hydrogen atom is replaced by deuterium. The Hessian of every bead is calculated by finite differences of the forces of the MLP, which reproduce the analytic harmonic frequencies of the MLP at the minimum to 0.2~cm$^{-1}$. With 1024 beads, the splitting is within about 1\% of its limit for infinitely many beads: on the MLP at $a_\mathrm{opt}$, 512, 1024 and 2048 beads give 29.16, 29.81 and 29.98~cm$^{-1}$ for H and 4.350, 4.395 and 4.406~cm$^{-1}$ for D; the steps shrink as the inverse square of the number of beads, which extrapolates to 0.8\% (H) and 0.4\% (D) above the 1024-bead values. The protocol checks are in Table~\ref{tab:si-coordinate-splitting-convergence}.

\begin{table}[h]
\caption{Convergence of the tunneling splitting (cm$^{-1}$) with the number of beads of the half ring polymer, on a test MLP at $a = 0.461$ (cc-pVDZ, one model), with the zero mode removed by projecting out the tangent of the path, and the action $S/\hbar$. All three runs are on one machine with the settings of the splittings of the main text: the minima optimized with the tightest criteria of geomeTRIC and polished by two Newton steps, i-PI's convergence thresholds tightened tenfold, and the masses of the most abundant isotopes. The splitting is 28.41, 29.33 and 29.57~cm$^{-1}$ with 512, 1024 and 2048 beads; the steps shrink as the inverse square of the number of beads, which extrapolates to 29.66~cm$^{-1}$, 1.1\% above its 1024-bead value. In pairs of runs on one machine, the splitting changes by $+3.2$\% from 25 to 12.5~K at 2048 beads, and at 512 beads by $-0.6$\% when i-PI's thresholds are tightened tenfold and by $-0.7$\% with the isotope masses instead of i-PI's standard atomic weights. At the minimum, the finite differences of the forces that give every Hessian of the calculation reproduce the analytic harmonic frequencies of the MLP to 0.16~cm$^{-1}$, which changes the zero-point energy by $-0.11$~cm$^{-1}$. The determinant with the zero mode projected out, from the block factorization used for every splitting, agrees with a dense calculation on the full Hessian of 3 runs to $3\times10^{-10}$ in its logarithm.}
\label{tab:si-coordinate-splitting-convergence}
\begin{tabular}{ccc}
\hline
beads & splitting & $S/\hbar$ \\
\hline
512 & 28.41 & 5.6747 \\
1024 & 29.33 & 5.6788 \\
2048 & 29.57 & 5.6793 \\
\hline
\end{tabular}
\end{table}

\paragraph{Results.}
With 1024 beads, ring-polymer instanton theory on the MLP gives tunneling splittings of 29.8~cm$^{-1}$ for H and 4.39~cm$^{-1}$ for D at $a_\mathrm{opt}$, 38\% and 51\% above the experimental 21.6 and 2.915~cm$^{-1}$.\cite{Baba1999_DetectionTunnelingRotation,Baughcum1984_MicrowaveSpectroscopicStudy} The five ensemble members give 28.3--31.8 and 4.14--4.77~cm$^{-1}$ (Figure~\ref{fig:coordinate}f; Table~\ref{tab:si-coordinate-splittings-dz}). The 512-bead values are 29.2 and 4.35~cm$^{-1}$, 2.2\% and 1.0\% below the 1024-bead results, and the 2048-bead values 0.6\% and 0.3\% above them. Across the displayed coordinate scan, the H splitting falls from 61.1~cm$^{-1}$ at $a=0.35$ to 12.0~cm$^{-1}$ at $a=0.60$.

\begin{table}[h]
\caption{Tunneling splittings (cm$^{-1}$) from ring-polymer instanton calculations on the final MLP (cc-pVDZ, 1024 beads, 25~K): the ensemble of five across the space and at $a_\mathrm{opt}$, and its five members alone at $a_\mathrm{opt}$; the barrier of the MLP (kcal/mol) and the action $S/\hbar$ of the instanton. With 2048 beads the splitting at $a_\mathrm{opt}$ is 29.98~cm$^{-1}$ (H, $+0.6$\% from 1024 beads) and 4.41~cm$^{-1}$ (D, $+0.3$\% from 1024 beads). $^\dagger$: below $a = 0.35$ the zero-point energy cancels the barrier of the MLP (at $a = 0.25$ it takes it from 2.7 to 0.2~kcal/mol), the proton is not localized and instanton theory, a theory of deep tunneling, does not apply; these rows are listed for completeness and are not shown in the main text.}
\label{tab:si-coordinate-splittings-dz}
\begin{tabular}{lccccc}
\hline
MLP & isotope & $a$ & barrier & $S/\hbar$ & splitting \\
\hline
ensemble & H & 0.200$^\dagger$ & 2.36 & 3.875 & 175.96 \\
ensemble & H & 0.250$^\dagger$ & 2.67 & 4.261 & 123.59 \\
ensemble & H & 0.300$^\dagger$ & 2.98 & 4.643 & 86.71 \\
ensemble & H & 0.350 & 3.29 & 5.016 & 61.06 \\
ensemble & H & 0.400 & 3.60 & 5.379 & 43.21 \\
ensemble & H & 0.450 & 3.91 & 5.728 & 30.89 \\
ensemble & H & 0.455 & 3.94 & 5.765 & 29.81 \\
ensemble & H & 0.500 & 4.22 & 6.064 & 22.27 \\
ensemble & H & 0.550 & 4.52 & 6.385 & 16.26 \\
ensemble & H & 0.600 & 4.82 & 6.694 & 11.98 \\
members, $a_\mathrm{opt}$ & H & 0.455 & 3.87--4.03 & -- & 28.28--31.81 \\
ensemble & D & 0.200$^\dagger$ & 2.36 & 4.925 & 41.75 \\
ensemble & D & 0.250$^\dagger$ & 2.67 & 5.412 & 26.54 \\
ensemble & D & 0.300$^\dagger$ & 2.98 & 5.893 & 16.91 \\
ensemble & D & 0.350 & 3.29 & 6.362 & 10.84 \\
ensemble & D & 0.400 & 3.60 & 6.819 & 7.01 \\
ensemble & D & 0.450 & 3.91 & 7.258 & 4.60 \\
ensemble & D & 0.455 & 3.94 & 7.305 & 4.39 \\
ensemble & D & 0.500 & 4.22 & 7.682 & 3.05 \\
ensemble & D & 0.550 & 4.52 & 8.086 & 2.06 \\
ensemble & D & 0.600 & 4.82 & 8.475 & 1.41 \\
members, $a_\mathrm{opt}$ & D & 0.455 & 3.87--4.03 & -- & 4.14--4.77 \\
\hline
\end{tabular}
\end{table}

The scan is restricted to fractions where the zero-point level lies well below the barrier, as required for this deep-tunneling instanton treatment. At B1LYP ($a=0.25$), the zero-point energy nearly cancels the barrier, reducing 2.7 to 0.2~kcal/mol; this is outside the regime where that treatment can reliably predict a splitting. At $a_\mathrm{opt}$, the zero-point level lies 1.3~kcal/mol below the barrier.

\paragraph{}

\paragraph{Cost.}
Each splitting uses about  2.5--2.6 million MLP energy-and-force evaluations. At the measured cc-pVDZ DFT timings, the corresponding DFT workload is about 25,000--32,000 core-hours, versus about 13 core-hours for the five-model MLP ensemble. All but about 55,000 of the MLP evaluations are in the three optimizations of the instanton, and 55,000 are for the finite-difference Hessian of the final instanton; a DFT energy and gradient with cc-pVDZ takes 34 (B1LYP alone, Intel Xeon Gold 6240 CPU) to 44~core-seconds (the five functionals in an earlier active-learning run used for this timing, on the same cluster), and a splitting on the ensemble of five MLPs, evaluated in double precision with the 64 beads of each i-PI batch in one call, takes 3.3~h on four cores of an Intel Xeon Gold 6240 CPU, timed on a compute node with no other jobs (no GPU is used). The MLP cost is the wall time multiplied by the four requested CPU threads (13.4 core-hours); the exclusive-node reservation accounts for 60.2 allocated core-hours, while Slurm records 10.7 hours of CPU use. These are workload estimates for this calculation, not a measured end-to-end DFT instanton run.

\FloatBarrier

\printsireferences
\end{document}